\documentclass[aps,pra,twocolumn,reprint,amsmath,amssymb,superscriptaddress,floatfix,footinbib,longbibliography]{revtex4-1}

\usepackage{epsfig}
\usepackage{graphicx}
\usepackage{epstopdf}

\usepackage[T1]{fontenc}
\usepackage[applemac]{inputenc}
\usepackage{lmodern}
\usepackage[english]{babel}

\usepackage{ae}
\usepackage{units}

\usepackage{amsmath,amssymb,natbib,bm}
\usepackage{psfrag}
\usepackage{subfigure}
\usepackage{amsthm}
\usepackage{bbm}

\usepackage{booktabs}

\usepackage[americaninductors]{circuitikz}
\usepackage{tikz}
\usetikzlibrary{arrows}

\usepackage{color}
\usepackage{url}

\usepackage[colorlinks]{hyperref}
\hypersetup{%
        plainpages=true,
        breaklinks=true,
        hypertexnames=false,
        pageanchor=true,
        colorlinks=true,
        linkcolor={blue},
        citecolor={magenta},
        urlcolor={blue},
        anchorcolor={black}
      }

\usepackage[all]{hypcap} 
      
\usepackage{mleftright} 

\usepackage{longtable} 

\usepackage{tabularx} 

\newcommand{\figref}[1]{\mbox{Fig.~\ref{#1}}}
\newcommand{\tabref}[1]{\mbox{Table~\ref{#1}}}
\newcommand{\secref}[1]{\mbox{Sec.~\ref{#1}}}

\renewcommand{\eqref}[1]{\mbox{Eq.~(\ref{#1})}}

\newcommand{\bra}[1]{\mleft\langle #1 \mright |}
\newcommand{\ket}[1]{\mleft|#1 \mright \rangle}
\newcommand{\braket}[2]{\langle #1|#2\rangle}
\newcommand{\ketbra}[2]{\mleft| #1 \rangle \langle #2 \mright|}

\newcommand{\expec}[1]{\mleft\langle #1 \mright\rangle}
\newcommand{\tr}[1]{\text{tr}\mleft( #1 \mright)}
\newcommand{\expecc}[1]{\text{E}\mleft[#1\mright]}

\newcommand{\comm}[2]{\mleft[ #1, #2 \mright]}
\newcommand{\lind}[1]{\mathcal{L}\mleft[#1\mright]}

\newcommand{\abs}[1]{\mleft|#1\mright|}

\newcommand{\abssq}[1]{\mleft| #1 \mright|^2}

\newcommand{\nn}{\nonumber}

\newcommand{\be}{\begin{equation}}
\newcommand{\ee}{\end{equation}}
\newcommand{\bea}{\begin{eqnarray}}
\newcommand{\eea}{\end{eqnarray}}

\newcommand{\rd}{\ensuremath{\mathrm{d}}}
\newcommand{\id}{\ensuremath{\,\rd}}
\newcommand{\figpanel}[2]{Fig.~\hyperref[#1]{\ref*{#1}(#2)}}
\newcommand{\figpanels}[3]{Fig.~\hyperref[#1]{\ref*{#1}(#2)-(#3)}}
\newcommand{\figpanelNoPrefix}[2]{\hyperref[#1]{\ref*{#1}(#2)}}

\AtBeginDocument{%
    \newwrite\bibnotes
    \def\bibnotesext{Notes.bib}
    \immediate\openout\bibnotes=\jobname\bibnotesext
    \immediate\write\bibnotes{@CONTROL{REVTEX41Control}}
    \immediate\write\bibnotes{@CONTROL{%
    apsrev41Control,author="08",editor="1",pages="0",title="0",year="1"}}
     \if@filesw
     \immediate\write\@auxout{\string\citation{apsrev41Control}}%
    \fi
}%

\begin{document}

\title{Classification and reconstruction of optical quantum states with deep neural networks}
\date{\today}

\author{Shahnawaz Ahmed}
\email{shahnawaz.ahmed95@gmail.com}
\affiliation{Department of Microtechnology and Nanoscience, Chalmers University of Technology, 412 96 Gothenburg, Sweden}

\author{Carlos S\'anchez Mu\~noz}
\affiliation{Departamento de Fisica Teorica de la Materia Condensada and Condensed Matter Physics Center (IFIMAC), Universidad Autonoma de Madrid, Madrid, Spain}

\author{Franco Nori}
\affiliation{Theoretical Quantum Physics Laboratory, RIKEN Cluster for Pioneering Research, Wako-shi, Saitama 351-0198, Japan}
\affiliation{Department of Physics, University of Michigan, Ann Arbor, Michigan 48109-1040, USA}

\author{Anton Frisk Kockum}
\email{anton.frisk.kockum@chalmers.se}
\affiliation{Department of Microtechnology and Nanoscience, Chalmers University of Technology, 412 96 Gothenburg, Sweden}

\begin{abstract}

We apply deep-neural-network-based techniques to quantum state classification and reconstruction. We demonstrate high classification accuracies and reconstruction fidelities, even in the presence of noise and with little data. Using optical quantum states as examples, we first demonstrate how convolutional neural networks (CNNs) can successfully classify several types of states distorted by, e.g., additive Gaussian noise or photon loss. We further show that a CNN trained on noisy inputs can learn to identify the most important regions in the data, which potentially can reduce the cost of tomography by guiding adaptive data collection. Secondly, we demonstrate reconstruction of quantum-state density matrices using neural networks that incorporate quantum-physics knowledge. The knowledge is implemented as custom neural-network layers that convert outputs from standard feedforward neural networks to valid descriptions of quantum states. Any standard feed-forward neural-network architecture can be adapted for quantum state tomography (QST) with our method. We present further demonstrations of our proposed~\cite{Ahmed2020} QST technique with conditional generative adversarial networks (QST-CGAN). We motivate our choice of a learnable loss function within an adversarial framework by demonstrating that the QST-CGAN outperforms, across a range of scenarios, generative networks trained with standard loss functions. For pure states with additive or convolutional Gaussian noise, the QST-CGAN is able to adapt to the noise and reconstruct the underlying state. The QST-CGAN reconstructs states using up to \textit{two orders of magnitude fewer iterative steps} than a standard iterative maximum likelihood (iMLE) method. Further, the QST-CGAN can reconstruct both pure and mixed states from \textit{two orders of magnitude fewer randomly chosen data points} than iMLE. Our work opens new possibilities to use state-of-the-art deep-learning methods for quantum state classification and reconstruction under various types of noise.

\end{abstract}

\maketitle

\tableofcontents


\section{Introduction}

Neural networks (NNs) are becoming ubiquitous in various areas of physics as a successful machine-learning (ML) technique to solve different tasks~\cite{Carleo2019}. Applications range from particle physics~\cite{Shlomi2020}, cosmology~\cite{Ravanbakhsh2017, Aragon-Calvo2019, Stein2020}, and many-body quantum matter~\cite{Carrasquilla2020} to material sciences~\cite{Schmidt2019}, and even to discover new physics~\cite{Iten2020, DAgnolo2019}. The NNs are used in classification problems, where the goal is to assign a label to a data sample~\cite{Zhang2000}, and for generative tasks, where new data is created after learning the underlying data distribution from samples~\cite{Kingma2014, Goodfellow2014}. Deep neural networks (DNNs) have shown impressive results in image classification~\cite{He2016, Chollet2017}, object detection~\cite{Redmon2015}, image denoising and inpainting~\cite{Xie2012, Zhang2017, Tian2020}, deconvolution~\cite{Xu2014}, generating realistic-looking images~\cite{Mirza2014, Isola2017, Karras2019, Mirsky2020}, text generation and translation~\cite{Brown2020, branwen2020gpt}, as well as for generating audio~\cite{Payne2019}, video~\cite{Suwajanakorn2017}, simulating gaming graphics~\cite{Kim2020}, and writing computer programs automatically~\cite{Gottschlich2018}. There are also recent examples of NN-based machine learning successfully applied to grand challenges in life sciences, e.g., the AlphaFold algorithm for protein folding~\cite{Senior2020, Alphafold2020}.

In quantum information and computing~\cite{Feynman1982, Nielsen2000, Montanaro2016, Wendin2017, Preskill2018, Arute2019}, some of the problems faced in characterizing and controlling quantum systems can be translated to tasks in ML. Many of these problems are data-driven and NN-inspired techniques have been used to successfully address them, e.g., identifying phase transitions~\cite{Rem2019}, detecting non-classicality or entanglement of quantum states~\cite{Sentis2015, Gebhart2020, You2020, Harney2020, Ma2018}, design of quantum experiments~\cite{Melnikov2018, ODriscoll2019}, quantum error correction~\cite{Torlai2017, Baireuther2017, Krastanov2017, Fosel2018, Fitzek2020}, characterizing and calibrating quantum devices~\cite{Flurin2020, Wittler2020} and foundational questions~\cite{Bharti2020}.

For quantum state characterization, even distinguishing two different quantum states can become challenging. For example, telling a coherent source of light and a thermal state apart can be difficult due to the close similarity of their the data in low-photon regimes~\cite{You2020}. Beyond just identifying properties of the quantum state or classifying them, reconstructing a full quantum state description presents an even more challenging task, called quantum state tomography (QST)~\cite{DAriano2003, Liu2005, Lvovsky2009, Ashhab2010}. The challenges arise mainly due to the exponentially large Hilbert-space dimension required to fully describe the state~\cite{Nielsen2000, Caves2004, Havlicek2019, Hou2016}. For example, $k$ two-level quantum systems (qubits) have a Hilbert space of dimension $N = 2^k$ and require up to $N^2 - 1$ real numbers to fully determine a density matrix describing the state. Therefore, QST requires clever data processing to extract a good representation of a state from noisy data~\cite{Deleglise2008, DAriano2009, Cramer2010, Flammia2011, Petz2012a, Baumgratz2013}. The presence of noise further complicates the problem; for additive Gaussian noise, one reconstruction method~\cite{Smolin2012} has computational complexity $O (N^4)$.

The success of NNs in other fields has prompted their application to several quantum state classification and reconstruction tasks. The motivation is that NNs are universal function approximators~\cite{Cybenko1989, Hornik1989, Schafer2007} that can learn maps from noisy input data to class labels, or act as variational ans\"atze for quantum states~\cite{Torlai2018, Glasser2018, Noe2019}. The variational ans\"atze can be learned from data by minimizing some loss metric between the predictions from the NN-based model and the data. From a computational learning perspective, approximately learning a quantum state has a linear scaling in the number of quantum bits~\cite{Rocchetto2019}.

\begin{figure*}[ht]
\centering
\includegraphics[width=\linewidth]{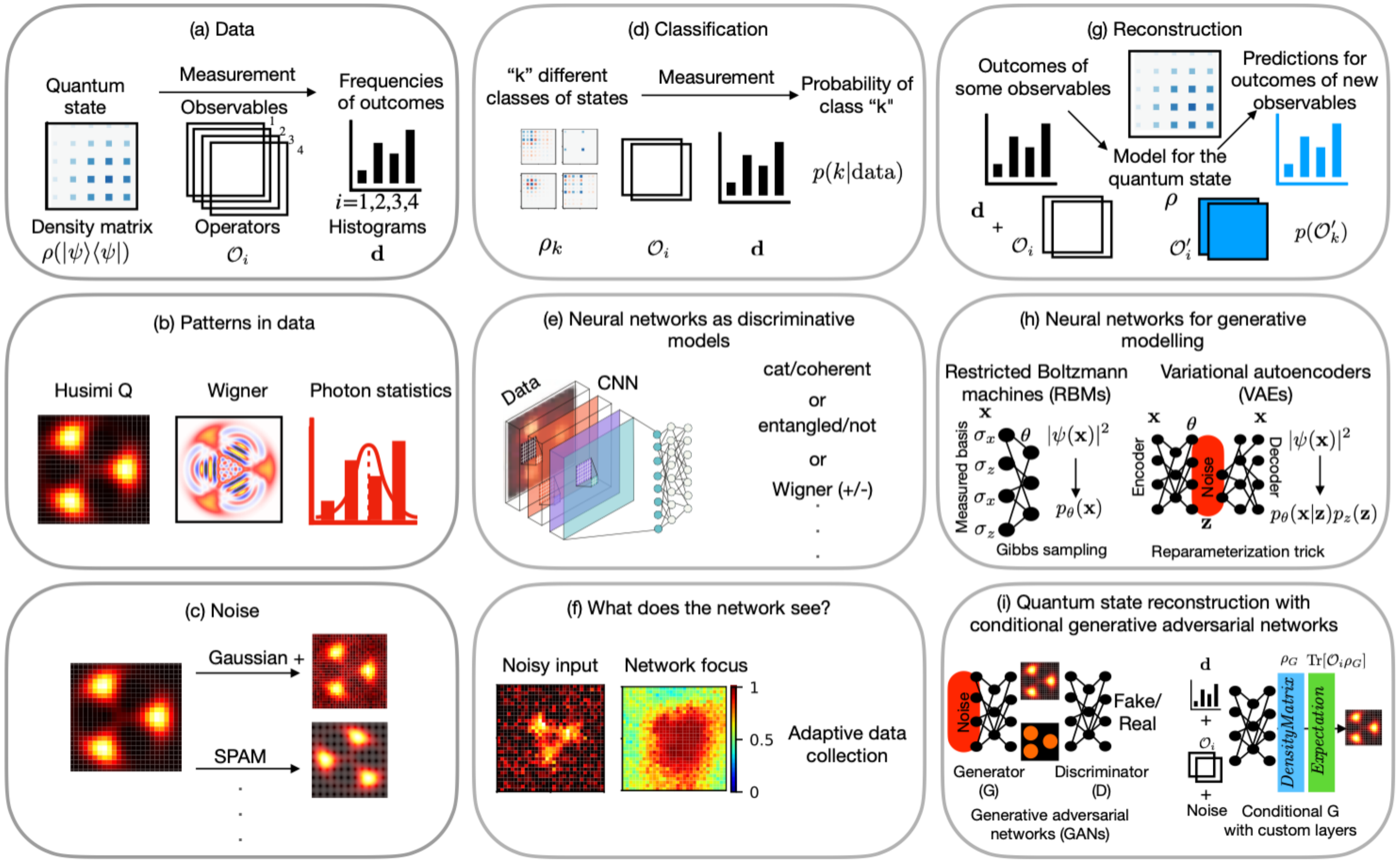}
\caption{An outline of the topics discussed in this work.
(a) The data required for quantum state tomography consists of the frequencies of measurement outcomes from observables represented as Hermitian operators. The aim is to reconstruct a description of the quantum state --- usually a density matrix or the wave function.
(b) Measurements of quasi-probability distributions (Wigner or Husimi $Q$) reveal interesting visual features in the data. Similarly, the histograms of measurement statistics, e.g., the photon-number distribution, can have patterns. Such features and patterns can be used for classification or reconstruction.
(c) Several types of noise can corrupt the state or the data. Some types of noise, e.g., white noise, can be reduced by more data collection. Other types of noise, e.g., state-preparation-and-measurement (SPAM) noise, are more difficult to handle.
(d) Classification tasks attempt to assign a label to data, classifying it according to its properties, e.g., if it is a Schr\"odinger-cat state, has Wigner negativity, or is an entangled quantum state.
(e) Neural networks can be trained for classification of states or their properties.
(f) Once it has been trained, analyzing how a neural network determines the class of a state can help to focus on the most important features in the data. This can be leveraged for adaptive data collection.
(g) Reconstruction of quantum states connects to generative modelling tasks, where the goal is to learn the underlying probability distribution of the data to sample new data from it. In quantum state reconstruction, we aim to learn an underlying model, usually in the form of a wave function or density matrix that can generate statistics for any measurement operator.
(h) Neural-network methods can be used for estimating or approximating underlying probability densities explicitly using restricted Boltzmann machines (RBMs) or variational autoencoders (VAEs). Training RBMs is not straightforward due to sampling requirements. This is resolved in VAEs using a reparameterization trick that allows gradient-based backpropagation for training.
(i) Generative adversarial networks (GANs) provide a density-estimation technique, where we do not explicitly define the density nor require the reparametrization trick for training. We combine ideas from VAEs and GANs to propose a new quantum state tomography technique with conditional GANs --- the QST-CGAN. Our QST-CGAN method allows for explicit estimation of the density matrix and computation of measurement statistics using two custom layers --- \textit{DensityMatrix} and \textit{Expectation}.
\label{fig:idea}}
\end{figure*}

In this work, outlined in \figref{fig:idea}, we connect the tasks of quantum state classification and reconstruction in a general way to discriminative and generative problems in ML. We demonstrate the feasibility of using DNNs for classification and reconstruction, showing how to flexibly adapt them for different scenarios, e.g., noise or scarce data. Crucial components of our methods include incorporating knowledge of quantum physics and other prior information into the network.

Many previous applications of DNNs for classifying quantum data~\cite{Gao2018, Gebhart2020, You2020, Harney2020}, consider properties like non-classicality or entanglement. In these works, more complicated noise models, beyond simple detection inefficiencies, are not considered. Since the classification task we tackle seems rather straightforward for DNNs, we attempt to go beyond the standard paradigm (training on simulated data, testing on new data) and demonstrate results with different types of noise for general states and measurements [see \figpanels{fig:idea}{a}{d}]. We also propose an adaptive data-collection method using a trained DNN to extract interesting patterns in the data and leverage it for adaptive tomography [see \figpanel{fig:idea}{f}].

In quantum state reconstruction [see \figpanel{fig:idea}{g}], one of the most popular neural-network approaches is to use Restricted Boltzmann Machines (RBMs) to map the underlying Boltzmann probability distribution of an RBM to the distribution of measurement outcomes on a quantum state~\cite{Torlai2018, Carleo2018a, Glasser2018, Tiunov2020} [see \figpanel{fig:idea}{h}]. This technique has some shortcomings, e.g., difficulties with sampling and lack of straightforward training for larger models. Recently, there has therefore been proposals to instead use feed-forward architectures, including recurrent neural networks (RNNs) and Transformers, for QST~\cite{Carrasquilla2019, Cha2020, Cai2018}. Unlike RBMs, such neural-network architectures are straightforward to train, without any need for sampling steps, using gradient-based optimization with backpropagation. However, state-of-the-art results for generative tasks in ML often use variational autoencoders (VAEs)~\cite{Kingma2014,Kingma2019} and generative adversarial networks (GANs)~\cite{Goodfellow2014, Mahdizadehaghdam2019}, which only recently are beginning to be explored for learning quantum states~\cite{Rocchetto2018, Zoufal2019, Hu2019, Lloyd2018} [see \figpanels{fig:idea}{h}{i}].

Results on the reconstruction of multi-qubit states suggest several benefits of NN-based reconstruction over standard techniques~\cite{Torlai2018a, Torlai2019, Lohani2020, Tiunov2020}. In Ref.~\cite{Cha2020}, states with up to 90 qubits are reconstructed in simulation. The ideas follow from Ref.~\cite{Carrasquilla2019}, where quantum state reconstruction using generative models, both RBMs and RNNs is combined with a tensor-network paradigm. Similarly, in Ref.~\cite{Palmieri2020}, fully connected DNNs were used for denoising data and dealing with state-preparation and measurement (SPAM) errors. In Ref.~\cite{Lohani2020}, a CNN was trained on simulated data (with noise) and proved able to reconstruct two-qubit states directly from data, outperforming a standard Stokes reconstruction.

However, such demonstrations are usually on simple states, with limited error models, and/or do not fully ensure that the reconstructed states are physical. For example, Ref.~\cite{Cha2020} considers Greenberger-Horne-Zeilinger (GHZ) states, which only contain four non-zero elements in the density matrix. In Refs.~\cite{Palmieri2020} and \cite{Lohani2020}, the Hilbert-space dimensions are restricted to six and four, respectively. Even then, techniques to include prior knowledge, such as the properties of quantum states or background noise, need to be explored. In Ref.~\cite{Lohani2020}, noise is handled by adding it to the simulated training dataset and properties of a quantum state are enforced using a similar idea to our proposal in Ref.~\cite{Ahmed2020} independently. In Ref.~\cite{Cha2020}, where GHZ states are reconstructed using a Transformer neural network, some reconstructed states have fidelities exceeding unity, which indicates the lack of quantum-mechanical constraints on the state description. In an experimental two-qubit reconstruction with the RBM ansatz, an improvement was observed when the variational ansatz was restricted to physical states, but this added costs during learning~\cite{Neugebauer2020}.

Furthermore, many of the approaches discussed so far cater specifically to qubit-based tomography. For continuous-variable (CV) quantum systems, which currently are attracting much attention for implementation of quantum computing~\cite{Lloyd1999, Gu2009, Weedbrook2012, Hillmann2020, Bourassa2020, Grimm2020, Joshi2020, Campagne-Ibarcq2020}, special adaptations are required, as in, e.g., Ref.~\cite{Tiunov2020}, where RBMs were adapted for CV systems, but required an exhaustive search of all possible configurations to train. Lastly, the reconstruction techniques usually either use DNNs to reconstruct a single state where data from one experiment is enough, or require training datasets~\cite{Lohani2020, Palmieri2020}. We show in Ref.~\cite{Ahmed2020} adaptions that allow the same DNN to both reconstruct states from scratch or perform single-shot reconstructions by mapping data to the space of density matrices in a general way.

Our motivation here is thus to address some of the problems discussed above and realize a unified framework that can flexibly work with different types of quantum data in various settings. Our contribution is a general method that allows the use of neural networks to capture patterns in data and explicitly generate a density-matrix description as an intermediate representation inside the network. This idea is inspired by developments in density estimation with neural networks~\cite{Papamakarios2019}, more specifically, the VAE architecture~\cite{Kingma2014}, which learns an underlying complex data distribution using a simple latent noise space to generate new data. We augment the latent space to be a full quantum state description (the density matrix) conditioned on inputs that are both the data samples and the operators that define the measurements. Using this conditioning allows us to have a very general technique that can further handle known noise --- we simply add the noise as an input variable. We consider the role of loss functions for reconstruction and motivate our idea of using a learnable loss in the form of a neural network based on the idea of GANs~\cite{Goodfellow2014, Mirza2014, Isola2017}. Our QST-CGAN technique thus combines concepts from VAEs and GANs, as illustrated in \figpanel{fig:idea}{i}. In this work, we present details of the implementation and results for noisy reconstruction, reconstruction of mixed states, and reconstruction from reduced data. 

This article is organized as follows. In \secref{sec:background}, we briefly discuss the quantum state tomography and state discrimination problems in the context of generative and discriminative modeling. In order to demonstrate our methods, we consider optical quantum states as examples. The various types of data from optical quantum states that will be used throughout the article are presented in \secref{sec:data}, including possible sources of noise. In \secref{sec:methods}, we describe details of the neural-network architectures and training methods. In particular, we discuss the custom layers that we introduce for reconstruction here and in Ref.~\cite{Ahmed2020}. Then, we present the results for the classification task in \secref{subsec:results-classification}, where we also analyse the impact of noise on the classification performance. In \secref{subsec:results-reconstruction}, we show the performance of the QST-CGAN on noisy data and the role played by various loss functions in the reconstruction. Finally, we conclude in \secref{sec:conclusion} and discuss, in \secref{sec:outlook}, further possibilities and potential for development of the techniques presented here. In \tabref{tab:abbreviations}, we list all the abbreviations used throughout the article for easy reference.

\begin{table}[]
\centering
\caption{List of abbreviations (in alphabetical order) used in this article.
\label{tab:abbreviations}}
\renewcommand{\arraystretch}{1.25}
\renewcommand{\tabcolsep}{0.15cm}
\begin{tabular}{ | m{5.8cm} | l |} 
\hline
\textbf{Full name} & \textbf{Abbreviation} \\
\hline
\hline
Compressed sensing & CS\\
\hline
Conditional generative adversarial network & CGAN\\
\hline
Continuous variable & CV \\
\hline
Convolutional neural networks & CNNs \\
\hline
Deep neural networks & DNNs \\
\hline
False positive rate & FPR\\
\hline
Generative adversarial networks & GANs \\
\hline
Gradient-weighted class activation mapping & Grad-CAM\\
\hline
Greenberger-Horne-Zeilinger & GHZ \\
\hline
Informationally complete & IC\\
\hline
Integral probability metrics & IPMs\\
\hline
Iterative maximum likelihood estimation & iMLE \\
\hline
Kullback-Leibler & KL\\
\hline
Machine learning  & ML \\
\hline
Matrix product state & MPS\\
\hline
Maximum likelihood estimation & MLE \\
\hline
Neural networks & NNs \\
\hline
Positive-operator-valued measures & POVMs\\
\hline
Projected gradient descent & PGD\\
\hline
Quantum state discrimination & QSD \\
\hline
Quantum state tomography & QST \\
\hline
Quantum state tomography with conditional generative adversarial network & QST-CGAN \\
\hline
Receiver-operating-characteristic & ROC\\
\hline
Recurrent neural networks & RNNs \\
\hline
Restricted Boltzmann machines & RBMs \\
\hline
State preparation and measurement & SPAM \\
\hline
Tensor network & TN\\
\hline
True positive rate & TPR\\
\hline
Variational autoencoders & VAEs \\
\hline
\end{tabular}
\end{table}


\section{Background}
\label{sec:background}

In this section, we set the stage for the paper by providing an overview of the problems of quantum state discrimination (QSD) and quantum state tomography (QST). We then discuss generative and discriminative modelling in machine learning, which is related to these problems. We compare different neural-network approaches to such modelling to motivate our choice of methods in this paper for tackling QST and QSD.


\subsection{Quantum state discrimination}

The task in QSD is to classify an unknown state $\rho$ as being one of a given finite ensemble of states $\{\rho_i\}$, from which states are chosen with probabilities $\{p_i\}$ such that $\sum_i p_i = 1$~\cite{Bae2015}. The classification is done by performing measurements on $\rho$, typically positive-operator-valued measures (POVMs) $\{ \mathcal O_i \}$, designed such that observing the outcome $i$, which occurs with probability $p_i = \tr{\mathcal O_i \rho}$, corresponds to the state being $\rho_i$.

The problem of QSD can thus often be rephrased as finding the optimal measurement for discriminating between the $\{\rho_i\}$. In case the states to be discriminated between are not orthogonal, perfect single-shot QSD is not possible. The optimal measurement should then instead maximize the probability of guessing the state correctly~\cite{Takagi2019}. Note that the non-orthogonality of quantum states, which prevents perfect QSD, does not have a classical analogue; it cannot be explained by merely assuming overlapping probability distributions~\cite{Bae2015, Leifer2013}.

If repeated state preparation and measurement is possible, adaptive measurement schemes, where new measurements are chosen based on the results of previous measurements, may be optimal. In this paper, we will consider such a situation, where we can make repeated measurements and collect statistics for various POVMs. However, our aim in this paper is not to construct highly optimized complex POVMs or adaptive schemes, but to show that a neural network can learn to perform QSD well when working with limited measurement data from standard, simple measurements of complex optical quantum states. Insights gleaned from the neural-network performance could then be used to minimize the number of simple measurements needed in experiments to classify states with high certainty. Furthermore, rapid state classification could help find a good starting point and parameterization for full quantum state reconstruction. Previous work has shown that neural networks can distinguish thermal and coherent light sources with few measurements~\cite{You2020}; here, we present a general framework for applying such techniques to arbitrary measurements and states. Note that we do not only distinguish between two types of states, but between many types of states at the same time. 


\subsection{Quantum state tomography}

The goal of QST is more ambitious than that of QSD: to fully characterize an unknown quantum state, usually by obtaining its density matrix $\rho$. A physical density matrix is Hermitian, positive semidefinite, and has unit trace. In an $N$-dimensional Hilbert space, $N^2 - 1$ real numbers have to be estimated from POVM outcomes to completely determine a general $\rho$. This can be seen clearly from the Cholesky decomposition
\be
\label{eq:cholesky}
\rho = T^\dag T,
\ee
which is extensively used in reconstruction methods to ensure positivity and Hermiticity. The matrix $T$ is lower-triangular with complex-valued entries except on the diagonal, where the entries are real-valued. 

The measurement data used for reconstruction of $\rho$ consists of single-shot outcomes from POVMs $\{ \mathcal O_i \}$. By repeating the measurement on identically prepared quantum states, we can gather statistics. The frequencies $d_i$ of various measurement outcomes is proportional to the expectation value $\tr{\mathcal O_i \rho}$ and forms our data $\mathbf d$. The reconstruction problem can therefore be stated as an inversion problem~\cite{Shen2016}
\be
\label{eq:sensing-matrix}
\mathbf d = A \rho_{f},
\ee
where the sensing matrix $A$ is given by the choice of measurement operators and $\rho_{f}$ is the flattened density matrix.

The invertibility of $\eqref{eq:sensing-matrix}$ depends on the set of measurement operators. A set of measurement operators that enables inversion, and thus allow the complete characterization of the state, is called informationally complete (IC)~\cite{DAriano2004}. For a state in an $N$-dimensional Hilbert space, up to $\sim N^2$ POVMs may be needed for IC (and each measurement needs to be repeated multiple times to gather the statistics). However, with some \textit{a priori} knowledge of the state, e.g., that $\rho$ is low rank or that certain elements of $\rho$ are zero, the measurements can be cleverly selected and their number reduced.

Reconstructing $\rho$ from $\mathbf d$ is thus an estimation problem, which can be approached in many ways. Common reconstruction techniques include linear inversion~\cite{Qi2013}, maximum likelihood estimation (MLE)~\cite{Hradil1997,Lvovsky2009,Banaszek2000}, and Bayesian methods~\cite{Blume-Kohout2010, Granade2016, Lukens2020}. Linear inversion, while being straightforward, can fail due to noise in the data or a high condition number of $A$~\cite{Shen2016} and produce unphysical entries in the density matrix, e.g., negative diagonal elements~\cite{Rehacek2007}. Therefore statistical inference techniques such as MLE or Bayesian estimation are preferred. Such methods give an estimate $\rho^{\prime}$ for the density matrix by optimizing the likelihood function
\be
L (\rho'|\mathbf d) = \prod_i \mleft[ \tr{\rho' \mathcal O_i} \mright]^{d_i}.
\label{eq:likelihood}
\ee
In case of continuous-variable outputs, where $d_i$ is a real number, appropriate binning is necessary to apply MLE~\cite{Silva2018}. Alternatively, the mean squared error between the output and the expected value can be minimized~\cite{Smolin2012}.

Although MLE guarantees a physical $\rho'$, it does not provide any error bars to quantify the uncertainty in the estimate. Recently, it has also been argued that MLE is not optimal and is an \textit{inadmissible} estimator for common metrics such as fidelity, mean-squared error and relative entropy~\cite{Ferrie2018}. Bayesian methods for QST, on the other hand, can quantify the uncertainity in the parameters of the density matrix using a prior probability distribution over different states $\pi(\rho)$~\cite{Blume-Kohout2010, Granade2016}. The initial prior $\pi_0(\rho)$ should be uniform, or as uninformative as possible, and is updated by applying the Bayes theorem using the likelihood $L (\rho' | \mathbf d)$ to give a posterior $\pi_f (\rho) \propto L (\rho | \mathbf d) \pi_0(\rho)$. The best estimate of the underlying state is given as the mean over all states $\rho_{\mu}$, defined by the posterior distribution $\pi_f$ weighted by the likelihood computed from observed data:
\be
\rho_\mu = \int \rho \pi_f(\rho) d \rho.
\ee

Other examples of methods to optimize the likelihood function and obtain a density matrix estimate include diluted MLE~\cite{Rehacek2007}, compressed sensing (CS)~\cite{Gross2010} and projected gradient descent~\cite{Bolduc2017}. The CS methods are motivated by simple parameter-counting arguments: we should only require $O (r N)$ measurements, with $r$ being the (low) rank of the density matrix~\cite{Kalev2015}. Examples of such low-rank states, common in experiments, are pure quantum states corrupted by local noise processes. Recently, other modifications of CS have been proposed and demonstrated experimentally for adaptive tomography~\cite{Ahn2019a, Ahn2019b}, which only require the a priori information of the density-matrix dimension (an improvement over CS, which requires an a priori guess of $r$).

However, a good ansatz or model for the state can reduce the effort for reconstruction. If we consider classes of quantum states having particular properties or symmetries, we can write their descriptions with fewer parameters than the $N^2 -1$ required for a general density matrix. Matrix-product-state (MPS)~\cite{Cramer2010,Lanyon2017a} and tensor-network (TN) tomography~\cite{Bridgeman2017,Chabuda2020} are methods that find efficient ans\"atze for states using MPSs or TNs, and permutationally invariant tomography~\cite{Toth2010,Moroder2012} exploits permutational symmetries of the density matrix. Some other improvements assume a noise model, e.g., additive gaussian noise~\cite{Smolin2012}, and therefore these techniques are often restricted to specific situations, lacking versatility.

A different formulation from the above techniques comes from the idea of projected gradient descent (PGD)~\cite{Bolduc2017}. In this method, a cost function is constructed that distinguishes between model-predicted data and the true data to apply gradient-based optimization to find the best estimate for the model (the density matrix). The benefit of the PGD technique is that it quickly converges to the MLE state in a wider variety of scenarios, even when the problem is ill-conditioned. The PGD method also sets up this notion of a cost-function, thereby translating the QST problem into an optimization problem.

Neural-network-based reconstruction methods have also shown significant promise. In such approaches, neural networks are either used as an ansatz for the state to obtain probabilities of measurement outcomes~\cite{Torlai2018a, Torlai2019, Tiunov2020}, or to directly estimate $\rho$~\cite{Lohani2020}. However, a general framework to study quantum state reconstruction using standard feedforward neural networks is missing. In this article, we present a framework that allows any standard neural network to be used for quantum state discrimination and reconstruction by adapting the generative and discriminative modelling framework from machine learning to QSD and QST.


\subsection{Discriminative and generative modeling}

Quantum state discrimination and reconstruction can be related to discriminative and generative tasks in machine learning. Consider a data space $\mathbf S$ from which we obtain samples $\mathbf x$ of a random variable $\mathbf X$. The samples can be classified as having one of $k$ different labels $y$. A data set can thus consist of a collection of pairs $\{\mathbf x, y \}$.

A discriminative model attempts to predict the class label $y$ for a data point $\mathbf x'$, i.e., finding the correct conditional probability $p( y | \mathbf x')$. We loosely interpret this as identifying whether a data point belongs to one of $k$ possible data distributions $p_{\texttt{data}}^{[k]}$.

A generative model aims to generate new samples $\mathbf x'$ that are similar to the observed data, which is assumed to be drawn from a data distribution defined by a probability density $p_{\texttt{data}}(\mathbf x)$. In general, real-world data distributions can be very complex, making it a hard problem to model them in a way that is both easy to compute and expressive enough to capture subtleties of the data. In \secref{sec:NNsGenDisc}, we discuss how deep neural networks are used to tackle such challenging distributions for generative and discriminative tasks.

The ideas of discriminative and generative modelling can be connected to QSD and QST in the following way. First, we identify the data space $\mathbf S$ with the space of measurement outcomes for operators $\{\mathcal O_i\}$. The outcomes can be collected either as single shots or average values; we denote the collected outcomes by $\mathbf d$. The expectation value $\expec{\mathcal O_i} = \tr{\mathcal O_i \rho}$ replaces the classical expectation value
\be
\expecc{\mathbf X} = \int \mathbf x p_{\texttt{data}}(\mathbf x) \, d \mathbf x.
\ee
Thus $\rho$ takes the role of a probability density function for the quantum system. If the data comes from one of $k$ different quantum states $\rho^{[k]}$, we can assign it a label $y$. Our data set is then formed by pairs $\{\mathbf d, y\}$.

The discrimination task of assigning one of the $k$ labels to some observed data $\mathbf d'$ is QSD. Reconstruction of $\rho$ can be considered a generative modelling task, where we aim to generate outcomes $\mathbf d'$ of new measurement operators $\{\mathcal O_i'\}$ after having observed some results of POVM measurements. To fulfil that task, we either need to obtain $\rho$ directly or find some parameterization of $\rho$ that lets us calculate $\expec{\mathcal O_i'} = \tr{\mathcal O_i' \rho}$.

Just like complicated classical data distributions $p_{\texttt{data}}$, $\rho$ can depend on many parameters and be difficult to estimate. However, efficient parameterizations of the quantum state using matrix-product states~\cite{Cramer2010, Lanyon2017}, tensor networks~\cite{Bridgeman2017, Chabuda2020}, or neural networks~\cite{Torlai2018, Carleo2018a, Glasser2018,  Cai2018, Carrasquilla2019, Cha2020, Tiunov2020, Palmieri2020, Lohani2020, Neugebauer2020} have reduced data and computation costs for quantum state reconstruction. In this article, we provide a general method to obtain $\rho$ as the output of neural networks, allowing the conversion of any neural-network architecture into a generative model for QST. Our ideas are applicable to any parameterization of $\rho$.


\subsection{Neural networks as discriminative and generative models}
\label{sec:NNsGenDisc}

Neural networks can approximate any function arbitrarily well~\cite{Cybenko1989}. They can be treated as functions that map an input space to a target space:
\be
f(\theta) : \mathbf S \to \mathbf T,
\ee
where $\theta$ are parameters that are learned from training on (labelled) data samples $\{\mathbf x, y\}$.

To use neural networks for discriminative tasks (classification) is fairly straightforward. In this case, the output $f(\mathbf x; \theta)$ of the network is interpreted as the conditional probability $p(y | \mathbf x)$~\cite{Zhang2000}. Then, by constructing a loss function that quantifies the total error of predictions on a training set, we can optimize the parameters $\theta$ to minimize the classification error.

Using neural networks as generative models is not as simple as mapping input data to target labels. Since a standard feed-forward neural network is a deterministic function $f(\mathbf x; \theta)$, it cannot be sampled to generate new data $\mathbf x'$. Early schemes used to circumvent this problem were neural networks with stochastic outputs, e.g., restricted Boltzmann machines (RBMs). Later, deterministic feedforward neural networks were adapted to give stochastic outputs for generative tasks; examples include variational autoencoders (VAEs) and generative adversarial networks (GANs). Below, we briefly discuss these methods to motivate our choice of using the conditional variant of GANs for quantum state reconstruction, and to show how our architecture also has connections to the other models.


\subsubsection{Restricted Boltzmann machines}

Restricted Boltzmann machines~\cite{Ackley1985, Smolensky1986, Hinton2012a} are stochastic neural networks that can represent arbitrary data distributions. An RBM consists of visible ($\mathbf v$) and hidden ($\mathbf h$) units which give stochastic binary outputs $\mathbf v, \mathbf h \in \{0, 1\}$. In single evaluations of the RBM, the states of the hidden units $h_j$ are updated to $1$ if the probability
\be
p (h_j = 1 | \mathbf v) = \sigma \mleft(b_j + \sum_i w_{i,j} v_i \mright)
\ee
is greater than a random number uniformly distributed between $0$ and $1$ (sampled in each update step). Here $\sigma$ is the sigmoid activation function and $\{b_j, w_{i,j}\}$ are parameters determining the interaction between different units. A visible unit $v_i$ is similarly updated depending on the states of the hidden units and another parameter $a_i$.

The result of updating the RBM units iteratively in this way from a random initial state is that the states of the visible units converge to a Boltzmann distribution
\be
p(\mathbf v;\theta) = \frac{1}{Z(\theta)} \sum_{\mathbf h} e^{-E(\mathbf v, \mathbf h; \theta)}, 
\ee
where $Z(\theta) = \sum_{\mathbf v, \mathbf h} e^{-E(\mathbf v, \mathbf h; \theta)}$ is the partition function and the energy is given by
\be
E(\mathbf v, \mathbf h; \theta) = - \sum_{i \in \text{visible}} a_i v_i - \sum_{j \in \text{hidden}} b_j h_j - \sum_{i,j} v_i h_j w_{ij},
\ee
parameterized by $\theta = {a, b, w}$. To train an RBM is to find parameters $\theta$ which make the probability distribution $p(\mathbf v;\theta)$ mimic the data distribution $p_{\texttt{data}}$, as measured by some statistical divergence, e.g., the Kullback--Leibler (KL) divergence. After training, new data points can then be generated by sampling $p(\mathbf v;\theta)$. Since standard RBMs only output binary-valued data, continuous-valued data needs to be handled either in a binary encoding or by using variants like Gaussian-Bernoulli RBMs~\cite{Krizhevsky2009}.

Although RBMs have been around for a long time, it was only recently that effective techniques for training them, e.g., contrastive divergence~\cite{Hinton2002, Hinton2012a}, were found and enabled them to play a significant role in the initial success of image processing with deep neural networks. These training methods have later been successfully applied to QST with RBMs~\cite{Carleo2017}. However, RBMs are still not straightforward to train and are less flexible than feedforward or convolutional neural networks. In particular, the partition function $Z(\theta)$ can be difficult to compute since it involves a sum over an exponential number of states~\cite{Manukian2020}. Furthermore, the sampling methods can have convergence issues for typical high-dimensional problems. These issues have stimulated the development of standard feedforward neural networks converted for generative modelling.


\subsubsection{Variational autoencoders}
\label{sec:VAE}

Variational autoencoders are an early example of an adaptation of standard feedforward neural networks to generative modelling. The idea of VAEs is to generate new data by sampling from a latent space $\mathbf Z$ and mapping it to the data space $\mathbf S$,
\be
\mathbf Z \xrightarrow{\texttt{Generator}} \mathbf S.
\ee
The latent space is used to define the data distribution, parameterized by $\theta$, as the marginal of a joint distribution $p_{\theta}(\mathbf x, \mathbf z)$ over the data and latent variables~\cite{Kingma2014, Ghosh2019, Kingma2019}:
\be
\label{eq:joint}
p_{\theta}(\mathbf x) = \int p_{\theta}(\mathbf x, \mathbf z) d\mathbf z.
\ee
The latent variable model $p_{\theta}(\mathbf x, \mathbf z)$ can be specified by using some prior noise distribution $p_{z}(\mathbf z)$, assuming the following factorization representing an infinite mixture model:
\be
\label{eq:vae_joint}
p_{\theta}(\mathbf x) = \int p_{z}(\mathbf z) p_{\theta}(\mathbf x|\mathbf z) d\mathbf z.
\ee
In VAEs, a neural network $g$ acts as a stochastic decoder to map the latent space to data:
\be
p_{\theta}(\mathbf x|\mathbf z) = p \mleft[ \mathbf S | g_{\theta}(\mathbf z) \mright].
\ee
Even if the factors in \eqref{eq:vae_joint} are simple, e.g., Gaussians, their mixture can be very expressive and thus capture complex data distributions.

However, the marginal $p_{\theta}(\mathbf x)$ is typically intractable due to the integral in \eqref{eq:joint}. Finding $\theta$ by some gradient-based optimization is thus not feasible. The intractable nature of the marginal stems from the intractability of the posterior
\be
p_{\theta}(\mathbf z|\mathbf x) = \frac{p_{\theta}(\mathbf x, \mathbf z)}{p_{\theta}(\mathbf x)}.
\ee
In VAEs, this posterior is approximated using a stochastic encoding in an encoder neural network $e$, parameterized by $\phi$, that maps the data space to the latent space:
\be
p_{\theta}(\mathbf z | \mathbf x) \approx q_{\phi}(\mathbf z|\mathbf x) = p \mleft[ \mathbf Z | e_{\phi}(\mathbf x) \mright].
\ee
The VAE architecture thus closely resembles that of an autoencoder -- a neural network that finds a compressed representation of data by encoding it in a latent space and reconstructing it back from there,
\be
\mathbf S \xrightarrow{\text{encoding}} \mathbf Z \xrightarrow{\text{decoding}} \mathbf X.
\ee

In general, VAEs assume $q_{\phi}(\mathbf z|\mathbf x)$ and $p_{\theta}(\mathbf x|\mathbf z)$ to be Gaussians specified by $g_{\theta}$ and $e_{\phi}$. The encoder network processes an input $\mathbf x$ to give the mean and covariance for a multi-dimensional Gaussian, which is sampled to obtain latent vectors $\mathbf z$. The decoder then generates new data $\mathbf x'$ from another multi-dimensional Gaussian with the mean and covariance determined by the sampled noise vector.

To obtain the parameters $(\theta, \phi)$, we want to maximize $\log {p_{\theta}(\mathbf x)}$, but since it is intractable, the variational approach maximizes the evidence lower bound or minimizes the loss
\be
E_{\mathbf x \sim p_{\texttt{data}}} \mleft[ -E_{\mathbf z \sim q_{\phi}(\mathbf z|\mathbf x)}\log{p_{\theta}(\mathbf x|\mathbf z)} + \text{KL}({q_{\theta}(\mathbf z|\mathbf x), p_{z}(\mathbf z)}) \mright].
\ee
However, training such a variational model comes with its own challenges due to the stochastic nature of the encoder and decoder. Even using a reparameterization trick making backpropagation-based training work on VAEs, several critical issues leave VAEs susceptible to generate samples that do not match the data distribution well. In image-generation tasks, this leads to blurry images as the Gaussian mixtures, used for their simplicity, are not the best for representing natural data distributions. 


\subsubsection{Generative adversarial networks}
\label{sec:GANs}

Generative adversarial networks~\cite{Goodfellow2014} and their conditional variant, conditional GANs (CGANs)~\cite{Isola2017}, solve the problem of approximating data distributions in a different way than RBMs and VAEs. In the GAN framework, a standard feedforward neural network $G$, with parameters $\theta_G$, generates new data using noise vectors $\mathbf z$:
\be
\mathbf x' = G(\mathbf z; \mathbf \theta_G). 
\ee
The network $G$ is trained by letting a second neural network, the discriminator $D$, evaluate the outputs from $G$. Unlike in a VAE, the second neural network does not map the input space to the latent space. Instead, the discriminator directly trains the generator to find the map from the latent noise space to data.

The discriminator $D$ is a standard classifier network, parameterized by $\theta_D$, that takes an input $\mathbf x'$ and outputs a probability $D(\mathbf x'; \mathbf \theta_D)$ that $\mathbf x'$ comes from the data distribution. The parameters $\{ \theta_G, \theta_D \}$ are optimized in an alternating fashion until the generator produces outputs that the discriminator cannot distinguish from samples of the real dataset, i.e., both $\mathbf x' \sim p_{\texttt{data}}$ and $\mathbf x \sim p_{\texttt{data}}$. In each optimization step, $\mathbf \theta_D$ is first updated to maximize
\be
E_{\mathbf x \sim p_{\texttt{data}}} [ \log(D(\mathbf x; \mathbf \theta_D))] + E_{\mathbf z \sim p_z} [\log (1 - D(G(\mathbf z; \mathbf \theta_G); \mathbf \theta_D))].
\label{eq:MaximizeD}
\ee
Then, $\mathbf \theta_G$ is updated to minimize
\be
E_{\mathbf z \sim p_z} [\log (1 - D(G(\mathbf z; \mathbf \theta_G); \mathbf \theta_D))].
\label{eq:MinimizeG}
\ee
When the training is completed, new samples can be generated from $G$ using noise vectors $\mathbf z$.

A standard GAN can only generate samples randomly according to the data distribution. However, we can modify the inputs to $G$ and $D$ by adding a conditioning variable $\mathbf c$ to guide the output. This leads to the CGAN architecture, where
\bea
G &:& {\mathbf z|\mathbf c} \to \mathbf x, \\
D &:& {\mathbf x|\mathbf c} \to \text{Pr}(\mathbf x \sim p_{\texttt{data}}).
\eea
The optimization of parameters for the CGAN networks follows the same procedure as for the GAN, i.e., maximizing \eqref{eq:MaximizeD} and minimizing \eqref{eq:MinimizeG}. 

The CGAN architecture provides a very flexible method for modelling complex conditional maps between different spaces. The flexibility stems from using the discriminator network, instead of a fixed loss function, as an evaluator of the generator performance. Conditional GANs have been used successfully in many types of generative tasks, e.g., generating images~\cite{Brock2018, Lucic2019}, converting edges to images, converting day to night pictures~\cite{Isola2017}, etc. In this article, we use CGANs for QST, but we also borrow ideas from VAEs for this task.


\section{Data}
\label{sec:data}

In this section, we define the data that we use for testing our methods of classification and reconstruction. We consider eight classes of optical quantum states, which we define in \secref{sec:states}. The data for these states is given by measurements consisting of applying coherent displacements followed by sampling of the photon number distribution for the resulting state, as we explain in \secref{sec:measurements}. We consider six types of noise, described in \secref{sec:noise}, that can distort the data.


\subsection{Optical quantum states}
\label{sec:states}

Optical quantum states are states of photons, i.e., of bosonic fields. In general, such states live in an infinite-dimensional Hilbert space, but we can obtain a finite-dimensional description by introducing a cutoff on the energy of the state. In the Fock basis for a single bosonic mode, a harmonic oscillator, the state is written
\be
\ket \psi = \sum_{n=0}^{N-1} c_n \ket{n}, 
\ee
where $n$ represents photon number, $N$ is the size of the Hilbert space, and $c_n$ are complex-valued amplitudes such that $\sum |c_n|^2 = 1$. Pure and mixed states in this Hilbert space are represented as $N \times N$ density matrices $\rho$.

Throughout this paper, we use a Hilbert-space cutoff of $N_c = 32$, except for some specific examples and demonstrations. We restrict the maximum photon number of the various states to $\lesssim 16$ to avoid artefacts due to truncation once the displacements are applied to these states. 

Below, we define the various types of states used in this work. The first three are well-known, basic classes of quantum optical states. The following four are from bosonic codes, i.e., states that are designed for quantum error correction. For these latter states, we adopt the definitions from Ref.~\cite{Albert2018}, where $\mu = \{0,1\}$ denotes whether the state encodes logical $0$ or $1$. Finally, we also use random states as noise for representing mixed states.


\subsubsection{Fock states}
\label{sec:fock}

The Fock states \texttt{fock} are the eigenstates of the Fock basis,
\be
\ket{\psi_{\texttt{fock}}} = \ket{n}.
\ee
We consider Fock states with photon number $1 \le n \le 16 $.


\subsubsection{Coherent states}

Coherent states \texttt{coherent} are displaced vacuum states, characterized by the complex displacement amplitude $\alpha$: 
\be
\ket{\psi_{\texttt{coherent}}(\alpha)} = \ket{\alpha} = D(\alpha) \ket{0},
\ee
where $D(\alpha) = \exp \mleft( \alpha a^\dagger - \alpha^* a \mright)$ is the displacement operator and $a$ ($a^\dag$) is the annihilation (creation) operator of the bosonic mode. The parameter $\alpha$ gives the position of the state in phase space. We consider $10^{-6}\le |\alpha| \le 3$ to keep the mean photon number $\abssq{\alpha}$ well below the Hilbert-space cutoff.


\subsubsection{Thermal states}

Thermal states \texttt{thermal} are mixed states where the photon number distribution follows super-Poissonian statistics:
\be
\rho_{\texttt{thermal}} (n_{\rm th}) = \sum_{n=0}^{N-1} p(n) \ketbra{n}{n},
\ee
where the probability distribution for the photons are given by
\be
p(n) = \frac{1}{n_{\rm th} + 1} \mleft(\frac{n_{\rm th}}{n_{\rm th} + 1} \mright)^n,
\ee
where $n_{\rm th}$ is the mean photon number. We consider thermal states with $n_{\rm th} \in [0, 16]$.


\subsubsection{\texttt{Num} states}

\texttt{Num} states are a specific set of bosonic-code states, consisting of superpositions of a few Fock states, numerically optimized (hence the name \texttt{num}) for quantum error correction, and characterized by their average photon number $\bar n \in \{ 1.562, 2.696, 2.770, 4.149, 4.336\}$~\cite{Michael2016, Albert2018}. The logical states for each code are orthogonal; for $\bar n =  1.562$, they are
\bea
\ket{\psi^{\mu=0}_{\texttt{num}} (1.562)} &=& \frac{1}{\sqrt{6}} \mleft (\sqrt{7 - \sqrt{17}} \ket{0} +  \sqrt{\sqrt{17}  - 1} \ket{3} \mright), \qquad \\
\ket{\psi^{\mu=1}_{\texttt{num}} (1.562)} &=& \frac{1}{\sqrt{6}} \mleft (\sqrt{9 - \sqrt{17}} \ket{1} +  \sqrt{\sqrt{17}  - 3} \ket{4} \mright). \qquad
\eea
%


\subsubsection{Binomial states}

Binomial states \texttt{bin} are bosonic-code states constructed from a superposition of Fock states weighted by the binomial coefficients~\cite{Michael2016, Albert2018}:
\be
\ket{\psi^{\mu}_{\texttt{bin}}} = \frac{1}{\sqrt{2^{N + 1}}} \sum_{m=0}^{N + 1} (-1)^{\mu m} \sqrt{ {N+1} \choose m} \ket {(S+1) m}.
\ee
Here the parameter $N$ plays a similar role as $\alpha$ in coherent states, determining the size of the state. For the parameter $S$, we use integers in the range $[1,10]$. Together with the Hilbert-space cutoff $N_c$, this determines a maximum value for N. We use $2 \leq N \leq N_c / (S+1) - 1$.


\subsubsection{Cat states}

Cat states \texttt{cat} are bosonic-code states consisting of superpositions of coherent states, with the simplest example being $(\ket \alpha \pm \ket {-\alpha})$ up to a normalization. In general, we can define cat states, parameterized by an integer $S$ and a complex displacement $\alpha$, as projections given by
\be
\ket{\psi^{\mu}_{\texttt{cat}}} = \frac{1}{\mathcal N} \Pi_{(S + 1)\mu}\ket {\alpha},
\ee
where $\mathcal N$ is a normalization. The projections are on even or odd Fock states given by 
\be
\Pi_r = \sum_{m=0}^{\infty} \ket{2 m (S + 1) + r} \bra{2 m (S + 1) + r},
\ee
with the variable $r \in \{ 0, 1, 2, ..., 2 S + 1\}$. The parameter $S$ corresponds to the number of photon-loss errors that the code can correct for. A simpler formulation for large $\alpha$, i.e., $2 \abs{\alpha} \sin ( \frac{\pi}{S + 1}) >> 1$, is an equal superposition of the $2(S +1)$ coherent states $\mleft \{ \ket {\alpha e^{i \frac{\pi}{S + 1}k}} \mright\}_{k=0}^{2S + 1}$ around a circle of radius $\abs{\alpha}$. We use $S \in \{ 0, 1, 2\}$ and $ \abs{\alpha} \in [1, 3]$.


\subsubsection{Gottesmann-Kitaev-Preskill states}
\label{sec:gkp}

Gottesmann-Kitaev-Preskill states \texttt{gkp} are bosonic-code states defined on a square grid in phase space~\cite{Gottesman2001, Albert2018, Campagne-Ibarcq2020}. The ideal \texttt{gkp} states can be seen as a superposition of vacuum states displaced to the points of this grid:
\be
\ket{\psi^{\mu}_{\texttt{gkp} (\text{ideal})}} = \sum_{{n_1, n_2} \in {\mathcal I}} \mathcal D \mleft( \sqrt{\frac{\pi}{2}} (2 n_1 + \mu) \mright) \mathcal D \mleft( i \sqrt{\frac{\pi}{2}}n_2 \mright) \ket 0
\ee
with the integers $n_1, n_2 \in \{-\infty, ..., -2, -1, 0, 1, 2, ..., \infty\}$ forming the grid.

However, a finite \texttt{gkp} state limits the lattice and adds a Gaussian envelope to make the state normalizable, thus parameterizing the state with a real parameter $\Delta \in [0, 1]$ as
\be
\ket{\psi^{\mu}_{\texttt{gkp} (\text{finite})}} = \sum_{\alpha \in \mathcal K(\mu)} e^{-\Delta^2 |\alpha|^2} e^{- i \rm{Re} [\alpha ] \rm{Im} [\alpha]} \ket \alpha,
\ee
where the complex amplitudes $\alpha$ are calculated from the grid $\mathcal K(\mu) = \sqrt{\frac{\pi}{2}} (2 n_1 + \mu)) + i \sqrt{\frac{\pi}{2}}n_2$ with some finite cutoff for $n_1, n_2$. We use $ n_1, n_2 \in \{-20, 20\}$ and $\Delta \in [0.2, 0.5]$. 


\subsubsection{Random states}
\label{sec:random}

Random states are mixed states generated using the QuTiP~\cite{Johansson2012, Johansson2013} function \texttt{rand\_dm} by choosing a density (proportion of non-zero elements) for the density matrix $\rho_\text{random}$. The elements of $\rho_\text{random}$ are sampled from a uniform distribution, ensuring that the density matrix is physical (Hermitian, positive semi-definite, and with unit trace). We choose the density $0.8$ for all tasks in this paper and allow $\rho_\text{random}$ to be full-rank mixed states.


\subsection{Measurements}
\label{sec:measurements}

Measurements on optical states are usually performed with a displace-and-measure technique. Applying a coherent displacement of amplitude $\beta$ and measuring the photon number distribution gives the generalized $Q$ function~\cite{Kirchmair2013}
\be
Q_n^{\beta} = \tr{\ketbra{n}{n} D(-\beta) \rho D^{\dagger}(-\beta)},
\ee
From the generalized $Q$ function we can easily obtain other quasi-probability distributions describing the state, e.g., the Husimi $Q$ function (photon field quadratures) 
\be
Q(\beta) = (1/\pi) Q^{\beta}_0
\ee
and the Wigner function (photon parity) 
\be
W(\beta) = (2/\pi) \sum_n (-1)^n Q_n^{\beta}.
\ee
%

\begin{figure*}[ht]
\centering
\includegraphics[width=\linewidth]{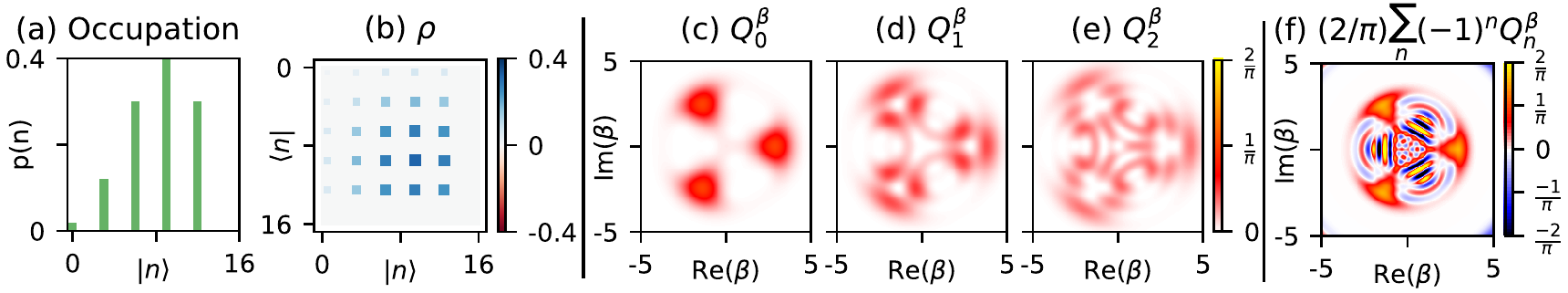}
\caption{A $\texttt{binomial}(S = 2, N = 5, \mu = 0)$ state and data generated from a displace-and-measure calculation using QuTiP~\cite{Johansson2012,Johansson2013} within a $200 \times 200$ grid.
(a) The photon occupation probabilities, i.e., the diagonal elements of the density matrix $\rho$.
(b) A Hinton plot of $\rho$, where blue (red) denotes that the real part of the density-matrix element is positive (negative). The size and the shade of each square is determined by the absolute value of the density-matrix element.
(c, d, e) The generalized $Q$ function, $Q_n^\beta$, for $n=0, 1, 2$.
(f) The corresponding Wigner function computed using the different $Q_n^\beta$ as $(2/\pi) \sum_n (-1)^n Q_n^{\beta}$. Note that even when choosing a Hilbert-space cutoff of $100$ for this demonstration, the corners in the Wigner-function plot have spurious non-zero values at large displacements $\beta \approx \pm 5 \pm 5i$. To mitigate such effects, larger cutoffs are required for states that have a high photon number or we need to restrict the computation to smaller values of $\beta$. Other methods of computing the Wigner function from $\rho$ do not suffer such problems even with a cutoff of $16$ for this specific example. QuTiP provides several such implementations and we use one of them, the numerically stable Clenshaw method, to compute Wigner functions in the rest of the paper.}
\label{fig:data}
\end{figure*}

In \figref{fig:data}, we plot the $Q_n^{\beta}$ functions for a \texttt{binomial} state to illustrate the different types of data. We also show how combining the various levels of the generalized $Q$ function leads to the Wigner function.  

In this paper, we mostly consider classification and reconstruction of optical quantum states based on Husimi-$Q$-function data, but our methods can also be used with Wigner-function data (as we show when reconstructing a state from experimental data in Ref.~\cite{Ahmed2020}), generalized-$Q$-function data, or data from any other observables.

\begin{figure*}[ht]
\centering
\includegraphics[width=\linewidth]{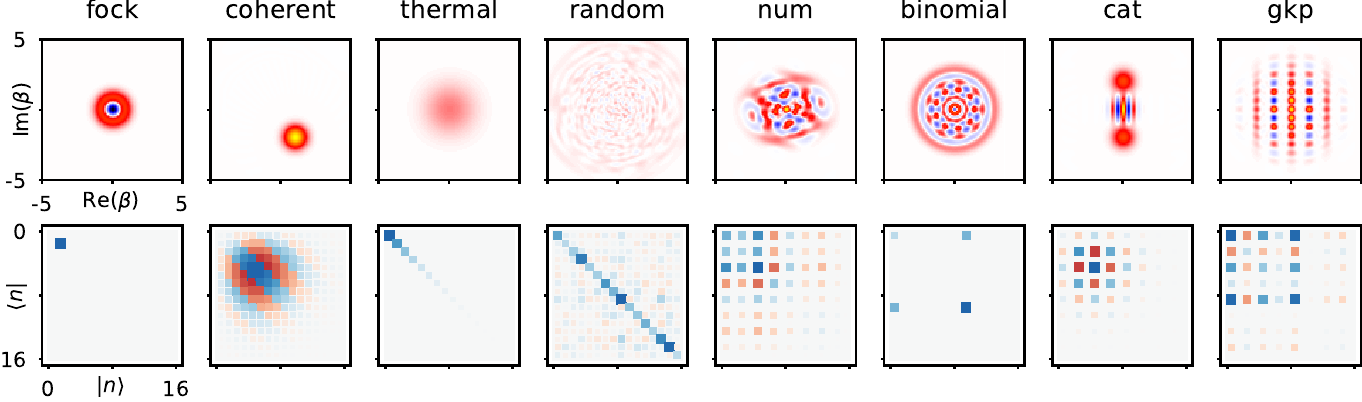}
\caption{Representative examples from each class of optical quantum states considered in this paper. In the top row, we plot the Wigner function for the states, using the same scaling as \figpanel{fig:data}{f}. In the bottom row, we show the values of the density-matrix elements for each state as Hinton plots similar to \figpanel{fig:data}{b}. We can see that the Wigner functions and density matrices have characteristic patterns that a neural network can learn and use for classification or reconstruction.}
\label{fig:zoo}
\end{figure*}

In \figref{fig:zoo}, we plot Wigner functions and Hinton plots of the density matrices for representative examples of all classes of states defined in \secref{sec:states} above.

\subsection{Noise}
\label{sec:noise}

Noise is an inevitable factor in most experiments. Thus, methods for state classification and reconstruction should be made sufficiently robust against various types of noise. In this subsection, we define the different types of noise that we use to test our neural-network-based classification and reconstruction.

Noise can enter the problem at different stages. First, the preparation of the state to be classified or reconstructed could have errors that lead to a slightly different state, $\rho \to \rho_\text{noisy}$ (state-preparation errors). Second, the measurement protocol could have errors due to calibration such that we are not measuring exactly what we sought out to measure, $\{\mathcal O_i \} \to \{ \mathcal {O}_i^\text{noisy}\}$ (measurement errors). Lastly, there can be errors in the data collection, e.g., errors incurred during amplification of the signal or photon shot noise which corrupts the data, $\mathbf d \to \mathbf {d}_\text{noisy}$ (data errors). 

The state-preparation and measurement (SPAM) errors can be systematic and thus hard to correct. Recently, deep neural networks have been demonstrated to be effective in learning such errors and correcting them~\cite{Palmieri2020} by training a supervised model to correct the data $\mathbf {d}_{\text{noisy}} \to \text{DNN} \to \mathbf d$. The neural network is thus used as a sophisticated filter to de-noise experimental data which can be agnostic to the underlying SPAM noise. In this work, during reconstruction, we do not train our networks to correct SPAM errors; we only deal with specific errors on a case-by-case basis. But for classification, we show that the neural-network approach is robust against the various types of SPAM and data errors defined below.

\begin{figure}[ht]
\centering
\includegraphics[width=\linewidth]{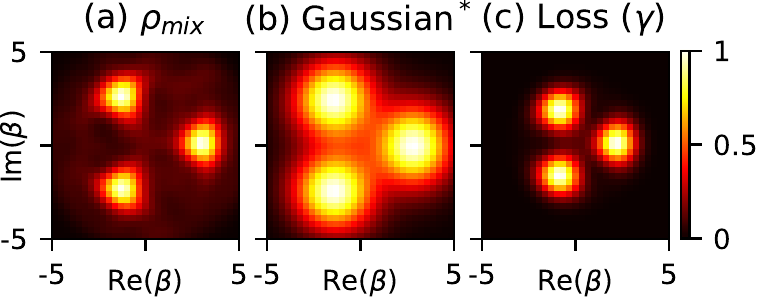}
\vspace{3mm}

\includegraphics[width=\linewidth]{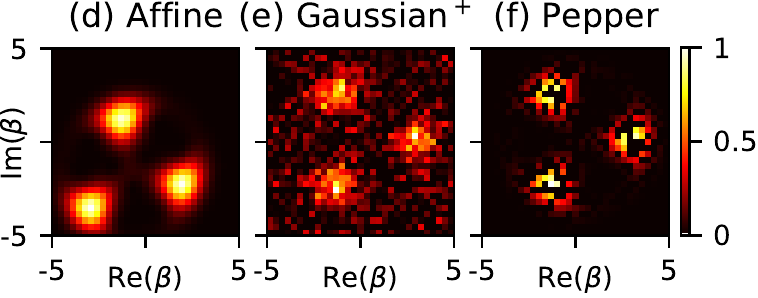}
\caption{The effect of various types of noise on the measurement data (Husimi $Q$ functions) for the state in \figref{fig:data}. We normalize the data to $[0, 1]$ by dividing with the maximum value and will use this color scheme throughout the paper to represent such rescaled data.
(a) Mixed states with $\sigma = 0.5$ and density 0.8 for $\rho_{\text{random}}$.
(b) Convolution with Gaussian noise with $n_{\rm th} = 3$.
(c) Photon loss with $\gamma \tau$ such that \unit[50]{\%} of the average initial photons have been lost.
(d) Affine transformation with rotation $\theta = 100^\text{o}$, shear $\Omega = 5^\text{o}$, and $\Delta x = \Delta p = 1.61$ ($5$ pixels).
(e) Additive Gaussian noise with standard deviation $\sigma_{G} = 0.2$.
(f) Pepper noise setting 50\% of the data points to zero.
}
\label{fig:noise}
\end{figure}


\subsubsection{Mixed states}
\label{subsubsec:mixed_noise}

In many experiments, thermal and other environmental noise will affect the quantum state. We model this noise by considering mixed states [see \figpanel{fig:noise}{a}]
\be
\rho_\text{mixed} = (1 - \sigma) \rho + \sigma \rho_\text{random},
\ee
with $\sigma \in [0, 0.5]$. In the classification task, the correct label for such a mixed state is defined to be that of the class that $\rho$ belongs to. In the reconstruction task, the aim would not be to reconstruct $\rho$, but to reconstruct $\rho_\text{mixed}$, since that is the actual state created in the experiment.


\subsubsection{Convolution with Gaussian noise during amplification}
\label{subsubsec:conv_noise}

In a measurement scheme which uses linear amplification detectors, one of the effects of noise is modelled by considering additional bosonic modes coming from the amplifier channel~\cite{Eichler2012}. The Husimi $Q$ function in the presence of such linear noise channels [see \figpanel{fig:noise}{b}] is a convolution
\be
Q_{\text{noisy}}(\beta') = \int_{\beta} P_h(\beta'^* - \beta^*) Q(\beta),
\ee
where $P_h(\beta^{\prime})$ is the Glauber-Sudarshan $P$ function~\cite{Drummond1980} of the noise mode. While at optical frequencies the noise mode is nearly in the vacuum state, such that $Q_{\text{noisy}}(\beta') \sim Q(\beta')$, at microwave frequencies, the noise mode is in a thermal state. In this case, 
\be
P_h(\beta) = \frac{1}{\pi n_{\rm th}} \exp \mleft(-\frac{|\beta|^2}{n_{\rm th}}\mright).
\ee
Therefore, the effect of noise is simply applying a Gaussian convolution with the variance $n_{\rm th}$. Note that such a noise is also interpreted as detection efficiency error with the reduced detection efficiency $\eta = 1/(1 + n_{\rm th})$. We consider such noise during reconstruction tasks by allowing it as an input which is easily estimated in experiments, e.g., the detector efficiency or thermal photons in the amplification channel.


\subsubsection{Photon loss}
\label{subsubsec:photon_loss_noise}

If the optical quantum state is created in a lossy resonator, photons may leak out from this resonator before the measurement of the state is completed. We model such photon loss [see \figpanel{fig:noise}{c}] by letting the original state evolve for some time $\tau$ according to the master equation
\be
\label{eq:master-eq}
\dot{\rho} = - \frac{i}{\hbar} \comm{H}{\rho} + \gamma \lind{a} \rho,
\ee
where $H = \hbar \omega a^\dag a$ is the free resonator Hamiltonian, $\omega$ is the resonator frequency, $\gamma$ is the photon loss rate, and $\lind{a} \rho = a \rho a^\dag - \frac{1}{2} a^\dag a \rho -\frac{1}{2} \rho a^\dag a$. Similar to the case of mixed states in \secref{subsubsec:mixed_noise}, in the classification task, the correct label is defined to be that of the class that $\rho (t = 0)$ belongs to, while in the reconstruction task, such a noise is not necessarily an error as the aim is to reconstruct $\rho (t = \tau)$.


\subsubsection{Affine transformations}
\label{subsubsec:affine_noise}

An affine transformation is a geometric transformation that can be represented as a composition of a linear transformation and a translation. In two-dimensional (2D) images, it preserves lines and parallelism, but allows for effects such as rotations, displacements, reflections, scaling, and shearing. Our motivation for this type of noise is that such effects can mimic SPAM errors, e.g., poorly calibrated displacement pulses, squeezing, and rotations of the state. We therefore consider rotations, displacements, scaling, and shearing to distort the training data (2D images of Husimi $Q$ or Wigner functions), see \figpanel{fig:noise}{d}.

If $(x, p)$ represent the position and momentum values in the phase space, i.e., $(\text{Re}(\beta), \text{Im}(\beta))$, the affine transformation $(x, p) \to (X, P)$ can be parameterized by the scaling factors ($s_x, s_y$), rotation angle $\theta$, the shear $\Omega$, and linear displacements $\Delta x, \Delta p$ as
\bea
X &=& s_x x \cos (\theta) - s_y p \sin (\theta + \Omega) + \Delta x, \\
Y &=& s_x x \sin (\theta) + s_y p \cos (\theta + \Omega) + \Delta p.
\eea
We use the TensorFlow~\cite{Tensorflow2015} implementation for data augmentation that applies such transformations, with the values of the parameters randomly selected within a certain range for each image augmentation: $\theta \in [0, 180^\text{o}]; \Omega \in [0, 5^\text{o}]; (\Delta x, \Delta p) \in [-2, 2]$ such that the pixels of the images are shifted up to \unit[20]{\%} of the image size. The range for scaling the image (zoom) is set to $0.2$ to allow shrinking or expanding the images within a factor $[0.8, 1.2]$ of the original size. We also allow the images to be flipped horizontally and vertically. The data augmentation described here is only used in the classification task.


\subsubsection{Additive Gaussian noise}
\label{subsubsec:additive_gaussian_noise}

Measuring the expectation value of a quantum observable often requires repeated measurements to find the average value with good precision. Thus, a limited number of measurements will reduce the precision. Moreover, the precision can also be reduced by binning of measurement results from nearby points in the phase space. We model these types of uncertainty in the data by adding randomly sampled values from a Gaussian distribution $\mathcal N$ with zero mean and standard deviation $\sigma_{G}$ to each data point as
\be
d_\text{noisy} = d + \mathcal N (0, \sigma_{G}).
\ee
See \figpanel{fig:noise}{e} for an example.


\subsubsection{Pepper noise}
\label{subsubsec:pepper_noise}

Salt-and-pepper noise represents a corruption of data where the signal changes drastically at a few points. We use pepper noise [see \figpanel{fig:noise}{f}] to represent dead pixels or missing data by selecting a random proportion of data points and setting them to zero.


\section{Methods}
\label{sec:methods}

In this section, we present the details of how we use deep neural networks for the two tasks -- classification (quantum state discrimination) and reconstruction (obtaining the density matrix) using the data discussed in \secref{sec:data}. Three different neural-network architectures are considered: \texttt{Classifier}, \texttt{Generator}, and \texttt{Discriminator}. We provide the methods and parameters for training and evaluation of the networks that we have used to obtain our results in this paper (\secref{sec:results}) and in Ref.~\cite{Ahmed2020}.


\subsection{Classification}

The problem of quantum state discrimination can be considered as a classification task, a task for which deep neural networks have shown impressive results. The input data $\mathbf d$ consists of observed frequencies for some measurement, which is related to the probabilities of outcomes of observables. The output is a label $\in \{\texttt{fock}, \texttt{coherent}, \texttt{thermal}, \texttt{cat}, \texttt{bin}, \texttt{num}, \texttt{gkp}\}$. The neural network we use for the classification is a standard convolutional neural network, which we train by minimizing cross-entropy loss using back-propagation.


\subsubsection{Input and output data}
\label{sec:ClassificationIO}

Since we consider optical quantum states, the data, e.g., the Husimi $Q$ or Wigner function of the state, can be rearranged into an image on a grid determined by the real and imaginary parts of the displacements $\beta$. Our training dataset is generated by randomly constructing states from the seven classes discussed in Secs.~\ref{sec:fock}-\ref{sec:gkp}, adding noise in the form of random mixed states (see Secs.~\ref{sec:random} and \ref{subsubsec:mixed_noise}) and then calculating the Husimi $Q$ functions of the resulting states for the fixed set of $\beta$ values evenly spaced in a $32 \times 32$ grid with $\beta \in [-5, 5]$.

We use 43,762 states for training and 8,670 states for testing. The input values are normalized to the range $[0, 1]$ by dividing each data instance with the maximum value. In the training phase, affine transformations (see \secref{subsubsec:affine_noise}) and additive Gaussian noise (see \secref{subsubsec:additive_gaussian_noise}) with $\sigma$ randomly selected between $[0, 0.05]$ are applied to the data. The addition of noise has a dual purpose - preventing overfitting and mimicing the effects of measurement noise. In the testing phase, we consider the impact of different types of noise separately.

The output labels are encoded in a $7$-dimensional vector using a one-hot-encoding, $\{t_i\}$ with $t_i \in \{0, 1\}$ and $t_i = 1$ denotes that the input state has been labelled as belonging to the class $i$. 

Note that the full generalized $Q$ function (see \secref{sec:measurements}) could be represented as a multi-channel image $n \times n \times N_c$, where $n \times n$ is the grid of $\beta$ values and $N_c$ is the photon-number cutoff. Similarly, we can just input the flattened data vector $\mathbf d$ for other types of measurements that cannot be seen as an image. However, for data in such form, using convolutional layers in the neural network would not make much sense, since there may not be any spatial correlations in the data.


\subsubsection{Network architecture}

\begin{figure}[ht]
\centering
\includegraphics[width=\linewidth]{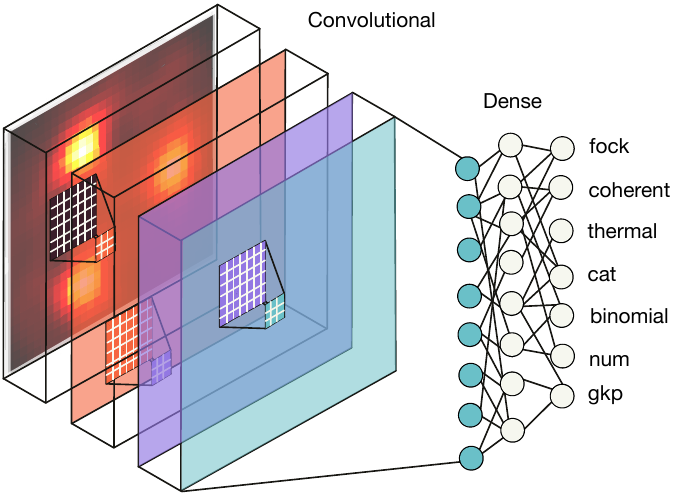}
\caption{Sketch of the \texttt{Classifier} network which classifies optical quantum states from Husimi $Q$ data (input image on the left). The blocks represents convolution operations, where filters (exemplified by boxes connecting one layer to another) extract features from the image. We use six such convolutional layers in our architecture (we only show three here). The extracted features are fed to the first of three fully connected layers. The outputs of the last layer are converted to a classification label. For the parameters used, see \tabref{tab:classifier}.
\label{fig:classifier}
}
\end{figure}

\begin{table}[ht]
\centering
\caption{Definitions, shapes, and number of trainable parameters for the layers of the \texttt{Classifier} network. We denote the convolution layers as $\text{Conv2D} (f, k, s)$ where $f$, $k$, and $s$ represent the filter size, kernel size, and strides respectively. After each convolution layer and the first dense layer, the activation function LeakyReLU is used. A full implementation of the code as a TensorFlow model can be found in Ref.~\cite{code}.
\label{tab:classifier}}
\renewcommand{\arraystretch}{1.25}
\renewcommand{\tabcolsep}{0.15cm}
\begin{tabular}{|l|c|r|}
\hline
Layer & Output shape & \# Parameters \\
\hline
Conv2D (32, 3, 1) & 30, 30, 32  & 288 \\
Conv2D (32, 3, 1) & 28, 28, 32  & 9,216 \\
Conv2D (32, 5, 2) & 14, 14, 32  & 9,216 \\
Conv2D (64, 3, 1) & 12, 12, 64 & 18,432 \\
Conv2D (64, 3, 1) & 10, 10, 64 & 36,864 \\
Conv2D (64, 5, 2) & 5, 5, 64 & 36,864 \\
Dense & 512 & 524,800\\
Dense & 256 & 131,328\\
Dense, output $y_i$ & 7 & 1,799\\
\hline
Total parameters & & 768,807 \\
\hline
\end{tabular}
\end{table}

The \texttt{Classifier} network, illustrated in \figref{fig:classifier} and detailed in \tabref{tab:classifier}, is a convolutional neural network (CNN). Its first six layers consists of blocks of convolution~\cite{LeCun1990} layers that extract geometric features from the input image. After the first six layers, the output is flattened and fed through two fully connected layers that output a 7-dimensional vector for each input image. 

We use the activation function ``LeakyReLU''~\cite{Maas2013} for all layers except the final output. The final output layer has 7 neurons, with outputs $\{y_i\}$, one for each class. We apply a softmax activation to these outputs,
\be
\text{softmax} (\mathbf y)_i = \frac{\exp(y_i)}{\sum_j y_j},
\label{eq:softmax}
\ee
to normalize the outputs such that they can be interpreted as the probability of the input data belonging to one of the seven classes. We assign the predicted label for the input state to the output that has the highest probability.  


\subsubsection{Training}

The parameters of the \texttt{Classifier} network are trained by minimizing the average cross-entropy loss between the predicted probabilities softmax($y_i$) in \eqref{eq:softmax} and the one-hot encoded target labels $t_i$, defined as
\be
\text{cross-entropy} (\mathbf t, \mathbf y) = - \sum_i t_i \log \mleft[ \text{softmax}(\mathbf y)_i \mright].
\ee
We use the gradient-based optimizer Adam~\cite{Kingma2014} with a learning rate $l = 0.0002$ and exponential decay rates for first and second moment estimates, $m_1 = 0.5, m_2 = 0.5$, to minimize the cross-entropy loss.

During training, we apply the dropout regularization technique~\cite{Srivastava2014}, where the output of a random fraction of neurons is ignored at each step of optimization, to prevent overfitting. We use \unit[40]{\%} dropout after the second, fourth, and sixth convolutional layers, and after the first dense layer. To further prevent overfitting, we also add a small (additive) Gaussian noise ($\sigma = 0.005$, see \secref{subsubsec:additive_gaussian_noise}) after the second and fourth convolutional layers.


\subsection{Reconstruction}

\begin{figure*}[ht]
\centering
\includegraphics[width=\linewidth]{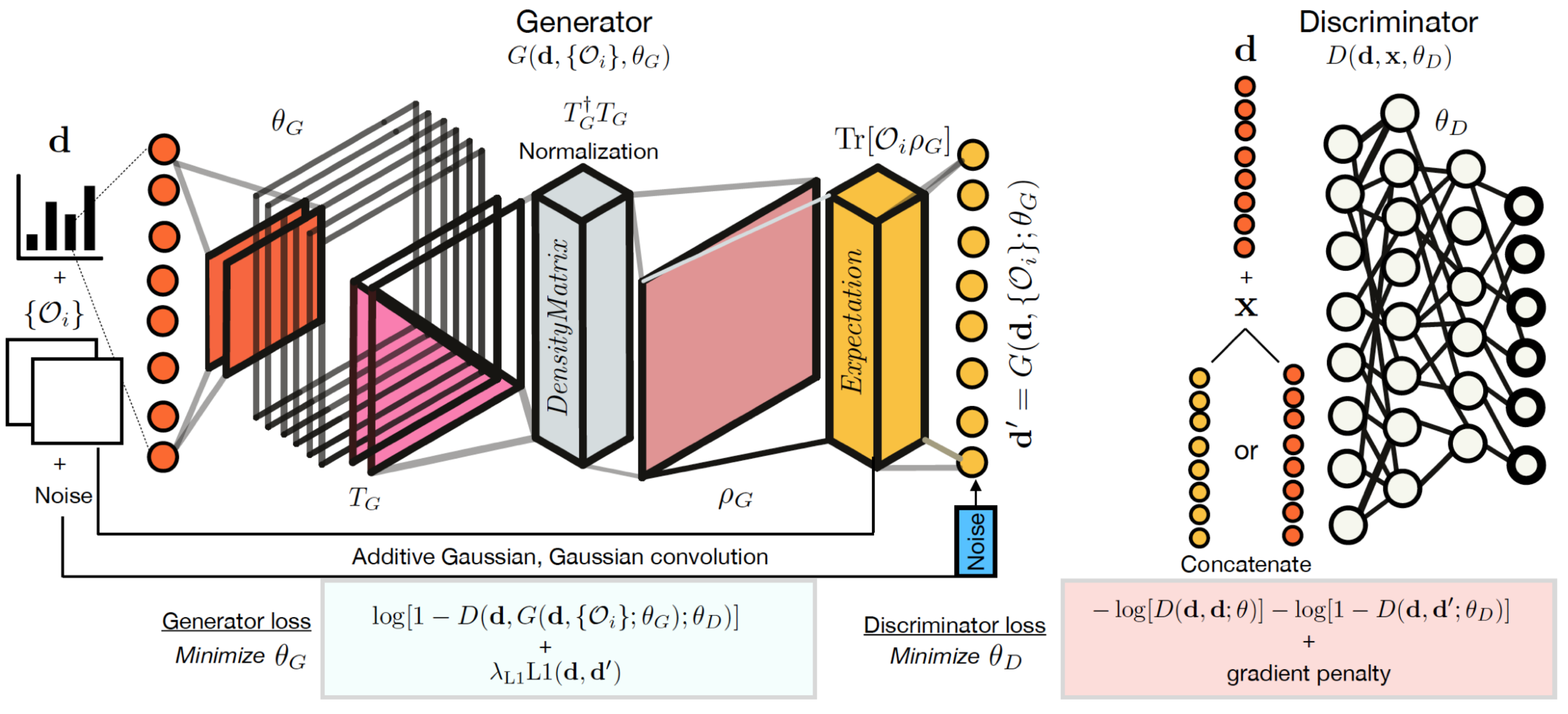}
\caption{Sketch of the \texttt{Generator} $G$ and \texttt{Discriminator} $D$ neural networks adapted for quantum state reconstruction. The two inputs to $G$ are the measurement statistics $\mathbf d$ and the observables $\{\mathcal O_i\}$. The input $\mathbf d$ is taken as a flattened vector, which is reshaped to a $16 \times 16 \times 2$ matrix after the first layer of $G$. Then, after successive transpose convolution operations, we obtain a $32 \times 32 \times 2$ matrix. This intermediate output is converted into a lower-triangular matrix with real elements on the diagonal to obtain a Cholesky decompostion form, $T_G$, that can yield a valid density-matrix representation $\rho_G$. The expectation values of the observables are then computed using the Born rule $\tr{\mathcal O_i \rho_G}$. In the last layer, any known source of noise is added to the outputs. The inputs to $D$ are a concatenation of $\mathbf d$ with either the generated data from $G$ or $\mathbf d$ itself. The output is interpreted as a similarity score between the inputs (the score is $\sim 1$ if they match, i.e., for inputs $\sim \mathbf d$). The weights of the two networks, $\theta_G$ and $\theta_D$ are updated alternatingly to minimize their respective loss functions. For the details of all the parameters, see \tabref{tab:generator} and \tabref{tab:discriminator}.
\label{fig:qst-cgan}
}
\end{figure*}

We now show how a standard neural network can be used to reconstruct the density matrix $\rho$ of a quantum state by adding custom layers to a generative model. The standard formulation of a generative model with feedforward neural networks (see \secref{sec:GANs}) is a map between a latent space and the data space. Our data $\mathbf d$ consists of single shots or average values of measurement outcomes for operators $\{\mathcal O_i\}$. We construct a \texttt{Generator} network parameterized by weights $\theta_G$,that first estimates a density matrix $\rho_G$. We then use a custom \textit{Expectation} layer that can generate the statistics for new measurements ${\mathbf d'}(d_i' = \tr{\mathcal O_i' \rho_G}$:
\be
\{ \mathbf d, \{\mathcal O_i\}\} \xrightarrow[G(\mathbf d, \{\mathcal O_i\}; \theta_G)]{\texttt{Generator}} \rho_G \xrightarrow[\tr{\mathcal O' \rho_G}]{\textit{Expectation}} \{\mathbf {d'} \}. 
\ee
The \texttt{Generator}-network formulation, depicted in \figref{fig:qst-cgan}, resembles a VAE (see \secref{sec:VAE}), but rather than modelling the data distribution using a parameterization with a mixture of Gaussians, we instead use the straightforward parameterization given by the estimated density matrix itself, $\rho_G$. The mapping between the latent space of measurement operators $\{\mathcal O_i'\}$ and the outcomes is simply $\tr{\mathcal O_i'\rho_G}$, which is the data generation map.

Below, we define the input and output data for the \texttt{Generator} network. Then, we show the details of the network architecture with our customized layers that regularize the intermediate output $\rho_G$ to a valid density matrix and generates the correct output. Finally, we discuss the training methods used to optimize the parameters of the \texttt{Generator} network. The first training method focuses on minimizing the least-squares and cross-entropy loss between the expected output and generated output. The second method learns a more sophisticated loss function in the form a second, trainable neural network, a \texttt{Discriminator}, also illustrated in \figref{fig:qst-cgan}. This second training method is inspired by the idea of CGANs~\cite{Isola2017, Goodfellow2014}, which we use for quantum state tomography (QST-CGAN)~\cite{Ahmed2020}. 


\subsubsection{Input and output data}
\label{subsubsec:reconst-data}

The input data for reconstruction are the measurement statistics $\mathbf d$ and the operators $\{\mathcal O_i\}$ that were measured. Similar to the classification task, we consider the Husimi $Q$ function in a $32 \times 32$ grid with $\beta \in [-5, 5]$. The measurement operators $\mathcal O_i$ are $32 \times 32$ complex-valued matrices. Therefore the input data for a single reconstruction is a combination of the flattened data vector $\mathbf d$ $(1 \times \text{1,024})$ and the set of operators $\{\mathcal O_i\}$ $(1 \times \text{1,024} \times 32 \times 32)$. Note that it is easy to change the parameters in the data or the neural-network architecture to allow arbitrary phase-space grid sizes and Hilbert-space cutoffs; the fact that they are both set to 32 in most examples here does not have any special significance.

The training data for a single reconstruction thus requires only these 1,024 data points (real-valued numbers) and the 1,024 operators (complex-valued matrices) as the input for each reconstruction. We consider noise on a case-by-case basis during training (described in \secref{sec:reconstruction-training} below).

The output of the neural network is a $(1 \times \text{1,024})$ vector representing the expectation values for the measurements $\{\mathcal O_i\}$. Inside the \texttt{Generator}, the full density matrix of the state is estimated as a $1 \times 32 \times 32$ complex-valued matrix $\rho_G$ determined by the outputs of an intermediate \textit{DensityMatrix} layer.

In this way, we allow for a flexible architecture, which can reconstruct a single state with inputs shaped as $(1 \times 1024, 1 \times 1024 \times 32 \times 32)$ for ($\mathbf d, \{\mathcal O_i\}$) or allow multiple states as the input simply by concatenating the inputs. For example, to reconstruct 10 states simultaneously with 1,024 measurements each, we simply feed the network a batch of data points as $(10 \times \text{1,024}, 10 \times \text{1,024} \times 32 \times 32)$.

In this article, we only consider single reconstructions, so our inputs will always be of the shape $1 \times n$ for the data $\mathbf d$ and $1 \times n \times N_c \times N_c$ for the measurements, where $n$ is the number of measurement settings and $N_c$ is the Hilbert-space cutoff. Note that we allowed the most general description of the measurement setting in the inputs as the full operator descriptions $\{\mathcal O_i\}$. We could also use alternative ways to specify the measurement settings, e.g., a set of complex displacements $\beta_i$, and redefine our \textit{Expectation} layer to use those $\beta$ values. In the case of qubit tomography, these measurement settings can be replaced with a set of single-qubit measurement operators such as $[Z, X, X, Z, \ldots ]$.


\subsubsection{Network architecture}
\label{subsubsec:cgan-architecture}

Our \texttt{Generator} network $G$ is a modified version of the standard $G(\mathbf z; \theta)$ formulation (see \secref{sec:GANs}), where we first consider the conditional form $G(\mathbf z|(\mathbf d, \{\mathcal O_i\}); \theta)$. The conditioning variable is our data and the measurement settings represented as a vector and a set of matrix operators, respectively. Then, inspired by the $pix2pix$ architecture~\cite{Isola2017}, we remove the random noise $\mathbf z$ and just consider the data and measurement operators as inputs to define $G(\mathbf d, \{\mathcal O_i\};\theta)$ as the $\texttt{Generator}$.

\begin{table}[]
\centering
\caption{Definitions, shapes, and number of trainable parameters for the layers of the \texttt{Generator} network. We denote the transpose convolution layers as $\text{Conv2D-T}(f, k, s)$ where $f$, $k$, and $s$ represent the filter size, kernel size, and strides respectively. After the first dense layer and the first three Conv2D-T layers, the activation function LeakyReLU is used. Instance normalization is used between the first two Conv2D-T layers. The output from the last Conv2D-T layer passes through two custom neural network layers: a \textit{DensityMatrix} layer generating $\rho_G$, and an \textit{Expectation} layer generating expectation values. A full implementation of the code as a TensorFlow model can be found in Ref.~\cite{code}.
\label{tab:generator}}
\renewcommand{\arraystretch}{1.25}
\renewcommand{\tabcolsep}{0.15cm}
\begin{tabular}{|l|c|r|}
\hline
Layer & Output shape & \# Parameters \\
\hline
Dense & 512 & 524,288 \\
Reshape & 16, 16, 2 & 0 \\
Conv2D-T (64, 4, 2) & 32, 32, 64 & 2,048 \\
Instance normalization & 32, 32, 64 & 128 \\
Conv2D-T (64, 4, 1) & 32, 32, 64 & 65,536 \\
Instance normalization & 32, 32, 64  & 128 \\
Conv2D-T (32, 4, 1) & 32, 32, 32 & 32,768 \\
Conv2D-T (2, 4, 1) & 32, 32, 2 & 1,024 \\
\textit{DensityMatrix} & 32, 32 & 0\\
\textit{Expectation} & 4,096 & 0\\
\hline
Total parameters & & 625,920 \\
\hline
\end{tabular}
\end{table}

The full architecture, detailed in \tabref{tab:generator} and depicted in \figref{fig:qst-cgan}, begins with a fully connected dense layer, which receives the flattened data vector $\mathbf d$ as input. The output of this layer is reshaped to a $16 \times 16 \times 2$ tensor. This layer converts the input into a matrix with two channels that can be upsampled into the density matrix. The next layers are three blocks of two-dimensional transpose convolution operations (Conv2D-T) and instance normalizations~\cite{Ulyanov2016} such that the final output is moulded to an estimate of the density matrix $\rho_G$. All the layers described so far use LeakyReLU activation, except the final Conv2D-T layer, whose outputs are fed to a custom \textit{DensityMatrix} layer.

The \textit{DensityMatrix} layer converts the output of the final Conv2D-T layer to a valid density matrix. This output is two matrices ($32 \times 32 \times 2$), which are combined into one $32 \times 32$ complex-valued matrix, $T_G$. The upper triangular part of $T_G$ and the imaginary part of the diagonal are set to zero to obtain the Cholesky decomposition of a Hermitian matrix [\eqref{eq:cholesky}]. Finally, we divide the resulting matrix by its trace to obtain a valid density matrix. Therefore, the custom \textit{DensityMatrix} layer can convert the real-valued outputs of any standard neural network to a Hermitian, positive-semidefinite matrix with unit trace.

The final layer is another custom one, called \textit{Expectation}. It takes as input $\{\mathcal O_i\}$ during training (the other part of the input to the $\texttt{Generator}$) and outputs the expected values for measurement outcomes for each component of $\mathbf d'$ as
\be
d_i' = \tr{\mathcal O_i \rho_G}.
\ee
The last two layers, \textit{DensityMatrix} and \textit{Expectation}, do not contain any trainable parameters.

\begin{table}
\centering
\caption{Definitions, shapes, and number of trainable parameters for the layers of the \texttt{Discriminator} network. The activation function LeakyReLU is used for all layers except the final output layer. A full implementation of the code as a TensorFlow model can be found in Ref.~\cite{code}.
\label{tab:discriminator}}
\renewcommand{\arraystretch}{1.25}
\renewcommand{\tabcolsep}{0.15cm}
\begin{tabular}{|l|c|r|}
\hline
Layer & Output shape & \# Parameters \\
\hline
Concatenate & 2,048 & 0\\
Dense & 128 & 1,048,704\\
Dense & 128 & 16,512\\
Dense & 64 & 8,256\\
Dense & 64 & 4,160\\
\hline
Total parameters & & 1,077,632\\
\hline
\end{tabular}
\end{table}

The \texttt{Discriminator} network used to train the generator is detailed in \tabref{tab:discriminator}. This network receives two inputs: the data $\mathbf d$ and the generated statistics $\mathbf d'$, and begins by concatenating the two. The concatenated input is then passed through four dense layers, with the final layer having $64$ neurons. All the layers of the discriminator use the LeakyReLU activation, except the final layer, whose outputs are interpreted as a measure of the similarity between $\mathbf d'$ and  $\mathbf d$. Note that the dimensions, or even the shape, of the final output layer can be arbitrary. The outputs simply need to be interpret as a similarity score between $\mathbf d$ and $\mathbf d'$ which should be $\sim 1$ if $\mathbf d \sim \mathbf d'$ and $\sim 0$ otherwise. We were inspired by the \textit{Patch}GAN idea~\cite{Isola2017} for our \texttt{Discriminator} that motivates penalties at the scale of patches in the input. We have also concurrently found during the course of our work, that similar ideas were effectively demonstrated for X-ray tomography with promising results for denoising~\cite{Liu2020a}.


\subsubsection{Training}
\label{sec:reconstruction-training}

The training for reconstruction can be done in two ways --- either we reconstruct a single state or we reconstruct a set of different states using the same $\texttt{Generator}$ network. This flexibility comes from our formulation of the $\texttt{Generator}$ network and reshaping of the data to find a map from data space to the set of density matrices. In this article, we only show how to perform single reconstructions, but in Ref.~\cite{Ahmed2020}, we show how the same $\texttt{Generator}$ network can perform single-shot reconstructions for many different states.

For each reconstruction in this article, we only consider a single state $\rho$ and the data from measurements of several operators on $\rho$ as the inputs and outputs (see \secref{subsubsec:reconst-data}). We train the $\texttt{Generator}$ network to minimize a loss metric that gives some measure of how the reconstructed statistics $\mathbf d'$, calculated from an underlying $\rho_G$, differ from the data $\mathbf d$. If $d_i$ are the frequencies of measurements $\mathcal{O}_i$ and $d_i' = \tr{\mathcal{O}_i \rho_G}$ are the computed probabilities from the generated density matrix, then maximizing the log-likelihood in \eqref{eq:likelihood} amounts to minimizing the cross-entropy loss between observed frequencies $\mathbf d$ and $\mathbf d'$:
\be
\text{cross-entropy}(\mathbf d, \mathbf d') = -\sum_i d_i \log [\tr{\mathcal {O}_i \rho_G}].
\label{eq:cross-entropy}
\ee
However, the cross-entropy loss assumes discrete-valued data, i.e., single-shot outputs of POVMs, whereas in many cases we may be looking at continuous-variable outputs instead. 

If we consider the data to be the expectation values of some continuous-valued observable, e.g., the homodyne current, metrics such as the mean squared error
\be
\text{L2} (\mathbf d, \mathbf d') = \frac{1}{q} \sum_i \mleft( d_i - d_i' \mright)^2,
\label{eq:L2}
\ee
where $q$ is the number of data points, are more suitable. For such continuous-valued data, the error in measurement can be assumed normally distributed with variance $\sigma_i^2$. Under this assumption, minimizing the L2 loss maximizes the likelihood
\be
L \mleft( \rho' | \mathbf d \mright) = \prod_i \mleft[ \frac{1}{\sqrt{2 \pi \sigma_i^2}} \exp \mleft( -\frac{\mleft( d_i - \mu_i \mright)^2}{2 \sigma_i^2} \mright)\mright],
\label{eq:likelihood-gaussian}
\ee
where we consider the mean for each measurement outcome as the expectation value $\mu_i = \tr{\rho' \mathcal M}$ for some observable $M$. 

Speaking more generally, the loss function uses some metric to measure the distance between two probability distributions $P$ and $Q$. Such metrics can be divided into two major classes: $\phi$-divergences and integral probability metrics (IPMs)~\cite{Sriperumbudur2009, Agrawal2020}. The first are of the form
\be
D_{\phi} (P || Q) = \int \phi \mleft( \frac{\partial P}{\partial Q} \mright) \id Q
\ee
where $\phi$ is some convex function $\phi: R_{\ge 0} \to R_{\ge 0}$, while the latter are defined as
\be
D_{\text{IPM}} (P, Q) = \sup\limits_{g \in \mathcal G} \abs{\int g \id P - \int g \id Q},
\ee
where the class of functions $\mathcal G$ parameterizes some notion of distance.

In the deep-learning community, the study of such metrics in generative modelling is an area of active research~\cite{Arjovsky2017,Bai2018}. It has been shown that the choice of loss function can greatly impact the quality of image reconstruction~\cite{Zhao2017}. There are several recent attempts to gain better understanding of the role of different loss functions in GAN performance, e.g., using the Wasserstein metric~\cite{Arjovsky2017} or IPMs~\cite{Sercu2017}. 

Since the best choice of loss function is far from clear, we train the \texttt{Generator} network to minimize several different loss functions between predicted Husimi $Q$ values and observed data. We first use the well-known L1, L2, and cross-entropy loss functions, as well as the KL divergence. The latter two are closely related, and belong to the class of $\phi$-divergences. The L1 loss is both a $\phi$-divergence and an IPM.

Beyond these well-known loss functions, GANs allow for more complex loss functions to be learned. In our QST-CGAN architecture, we train the $\texttt{Generator} $ using the \texttt{Discriminator} network combined with L1 loss, to minimize
%
\bea
&&\log \mleft[ 1 - D(\mathbf d, G(\mathbf d, \{\mathcal O_i\}; \mathbf \theta_G); \mathbf \theta_D) \mright] \nn \\
&+& \lambda_{\rm L1} \mleft[ G(\mathbf d, \{\mathcal O_i\}; \mathbf \theta_G) - \mathbf d \mright],
\label{eq:DiscriminatorL1}
\eea
with the L1 loss coefficient $\lambda_{\rm L1} \in \{ 0, 1, 10, 100 \}$. The discriminator loss function maximizes \eqref{eq:MaximizeD} by minimizing
\bea
&-& \log \mleft[D(\mathbf d, \mathbf d; \theta_D) \mright] - \log \mleft[ 1 - D(\mathbf d, \mathbf d'; \theta_D) \mright] \nn \\ 
&+& \lambda_{\Delta} E \mleft[ \mleft( || \Delta_{\mathbf x} D(\mathbf x; \theta_D) ||_2 - 1 \mright)^2 \mright],
\eea
where the last term is a gradient penalty~\cite{Gulrajani2017} with weight $\lambda_{\Delta} = 10$. We combined the inputs to the $\texttt{Discriminator}$ as the vector $\mathbf x$.

Therefore, in each training iteration, we alternatively update the generator and discriminator weights using backpropagation with the help of some gradient-based optimizer. Since the choice of hyperparameters, e.g., optimizer or learning rate, can significantly affect the rate of convergence, we try to find settings that enable a fast convergence for all loss functions. To make a fair comparison, we keep the same parameters for optimization for all loss functions. We use the Adam optimizer with the exponential decay rates for first and second moment estimates as $m_1 = 0.5, m_2 = 0.5$. We also use an exponentially decaying learning rate as a function of iteration number $i$, 
\be
l (i) = l_0 C^{\frac{i}{s}}
\ee
with initial learning rate $l_0 = 0.0002$, decay constant $C = 0.96$, and $s =1000$ steps.


\section{Results}
\label{sec:results}

In this section, we characterize the performance of our \texttt{Classifier} and \texttt{Generator} networks in various settings. We first check the performance of the \texttt{Classifier} network, including some types of noise, in \secref{sec:confusion_matrix}. We then study, in Secs.~\ref{sec:recognizing_cats_photon_loss} and \ref{subsub:results-gaussian-noise}, the impact of photon loss and additive Gaussian noise on classification performance. Finally, in \secref{sec:GradCAM}, we analyze which parts of the data the \texttt{Classifier} bases its decision on. This provides information that can help reduce the number of measurements needed in an experiment or guide an adaptive scheme for tomography.

For reconstruction, we first investigate, in \secref{sec:loss-functions}, the result of using different loss functions, including the \texttt{Discriminator} network, to train the \texttt{Generator} network. We show the performance of the \texttt{Generator} network against a standard maximum-likelihood-based reconstruction algorithm (iMLE)~\cite{Lvovsky2004} under additive Gaussian noise in \secref{sec:additive-gaussian-results} and gaussian convolution noise in \secref{sec:conv-gaussian-results}. Then, we show the results of reconstruction for mixed states in \secref{sec:mixed-states-results} and finally, in \secref{sec:data-reduction-results}, demonstrate how few data points are needed for the network to reconstruct a state well.


\subsection{Classification}
\label{subsec:results-classification}


\subsubsection{Confusion matrix}
\label{sec:confusion_matrix}

\begin{figure}[ht]
\centering
\includegraphics[width=\linewidth]{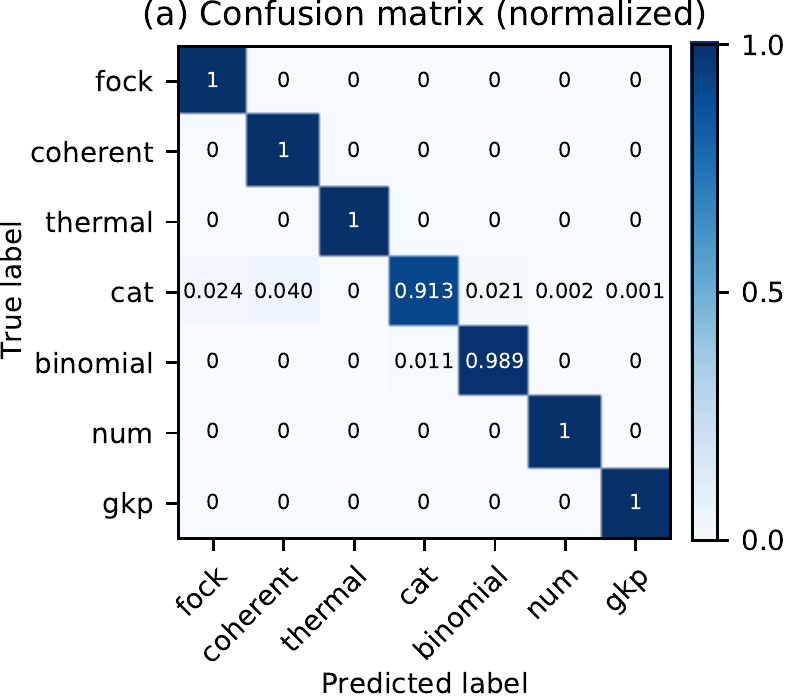}
\vspace{0mm}

\includegraphics[width=0.98\linewidth]{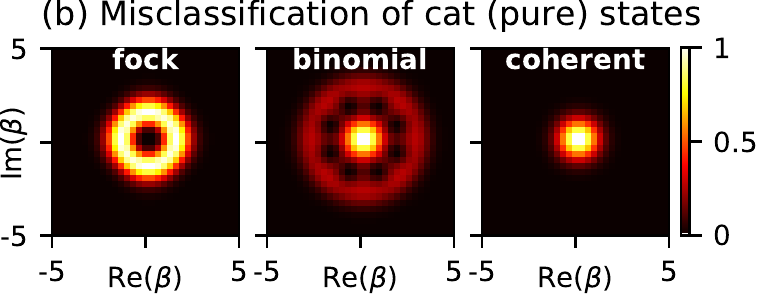}
\caption{Performance of the $\texttt{Classifier}$.
(a) Confusion matrix demonstrating the performance of the $\texttt{Classifier}$ on a test dataset containing $8,760$ states ($\sim 1,200$ different instances of each class). The prediction counts are normalized to show the true labels versus the predictions made by the $\texttt{Classifier}$.
(b) Husimi $Q$ functions for three examples of pure \texttt{cat} states that the $\texttt{Classifier}$ does not classify correctly. For each state, the incorrectly assigned label is shown. The states are, from left to right, $\texttt{cat}(\alpha = 1, S = 1, \mu = 1)$, $\texttt{cat}(\alpha = 2, S = 3, \mu = 0)$, and $\texttt{cat}(\alpha = 1, S = 3, \mu = 0)$. For certain parameters, different states have a high overlap in fidelities and the measurement data, making classification challenging. The $\texttt{Classifier}$ tries to find the best label according to relevant patterns in the data.
\label{fig:confusion}}
\end{figure}

The performance of the $\texttt{Classifier}$ network on a test set is shown as a confusion matrix in \figpanel{fig:confusion}{a}. The test set consists of $\sim 1,200$ different instances of each of the seven classes in Secs.~\ref{sec:fock}-\ref{sec:gkp}, with noise in the form of state mixing (see \secref{subsubsec:mixed_noise}) applied with $\sigma \in [0, 0.5]$ and density $0.8$.

The accuracy of the classification (number of correct classifications divided by the total number of classifications) on the whole test set is \unit[98.6]{\%}. For a validation set with the same states as the test set, but where we have added noise in the form of affine transformations (see \secref{subsubsec:affine_noise}) and additive Gaussian (see \secref{subsubsec:affine_noise} with $\sigma \in [0, 0.05]$) on top of the state-mixing noise, the accuracy of the $\texttt{Classifier}$ remains very high, \unit[97.7]{\%}.

It is clear from the confusion matrix in \figpanel{fig:confusion}{a} that the class which presents challenges for the network is \texttt{cat}. All other classes are correctly identified in virtually every case, but the \texttt{cat} states are misclassified in about \unit[9]{\%} of the cases. In these cases, the network misidentifies the \texttt{cat} states as all other classes except \texttt{thermal}, with the most common mislabellings being \texttt{coherent}, \texttt{fock}, and \texttt{binomial}. The reverse misidentification, where a state is misclassified as \texttt{cat}, occurs for about \unit[1]{\%} of the \texttt{binomial} states.

A few examples of misclassifications are shown in \figpanel{fig:confusion}{b}, where we consider pure $\texttt{cat}$ states with low values of $\alpha$. These examples demonstrate that there are parameters for which states from different classes are very similar. For example, a $\texttt{cat} (\alpha = 4, S = 0)$ state $\rho$ and a $\texttt{binomial} (S = 1, N = 16)$ state $\rho'$ have a fidelity
\be
F (\rho, \rho') = {\left[\tr{\sqrt{\sqrt{\rho} \rho' \sqrt{\rho}}}\right]}^2
\ee
greater than $0.99$. Note that the fidelity $F$ reduces to the squared overlap $\abssq{\braket{\psi}{\psi'}}$ for pure states. Similarly, the fidelity of $\texttt{cat} (\alpha = 3, S = 4)$ and $\texttt{fock} (10)$ is greater than $0.996$. It is thus not surprising that the network found some states hard to classify. A human quantum physicist would likely have made the same misclassifications from the data in \figpanel{fig:confusion}{b}.


\subsubsection{Recognizing cat states with photon loss}
\label{sec:recognizing_cats_photon_loss}

\begin{figure}[ht]
\centering
\includegraphics[width=0.98\linewidth]{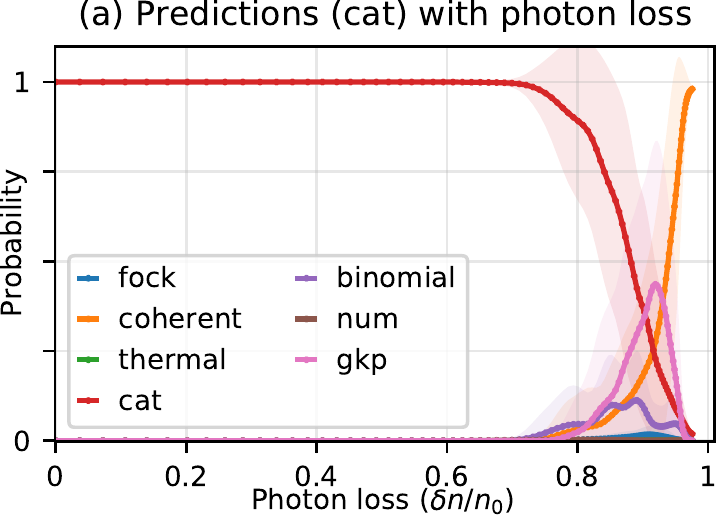}
\vspace{2mm}

\includegraphics[width=0.98\linewidth]{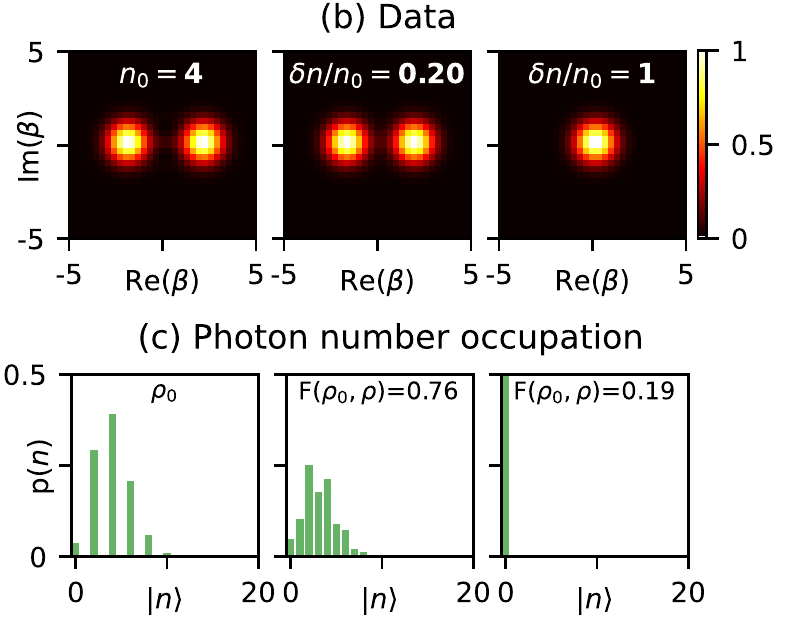}
\caption{$\texttt{Classifier}$ performance for \texttt{cat} states with photon loss.
(a) Softmax probabilities (solid lines) predicted by the $\texttt{Classifier}$ for the labels of the seven classes. The shaded regions show one standard deviation from the mean. The dataset consists of $100$ \texttt{cat} states ($\abs{\alpha} \in [2, 3], S = 0, \mu = 0$) with photon loss, quantified by the proportion of photons lost $\delta n / n_0$ starting from the initial mean photon number $n_0$.
(b) Husimi $Q$ functions for one of the \texttt{cat} states in the dataset with, from left to right, $\unit[0]{\%}$, $\unit[20]{\%}$, and $\unit[100]{\%}$ of photons lost with respective fidelities $0.76$ and $0.19$ for the states with photon loss. It is not straightforward to assert from just the Husimi-$Q$ data when a $\texttt{cat}$ state stops being a ``\texttt{cat}" as it still possesses $\texttt{cat}$-like features (two coherent blobs) even after losing $\unit[20]{\%}$ of the initial photons.
(c) The photon-number occupation probabilities for the states in (b). Note that the occupation probability for the vacuum state is $\sim 1$ but we set the limits of the y-axis to $0.5$ for better distinguishability.
}
\label{fig:photon-loss}
\end{figure}

We now investigate the performance of the \texttt{Classifier} network in the presence of photon loss (see \secref{subsubsec:photon_loss_noise}). In \figpanel{fig:photon-loss}{a}, we show how well the \texttt{Classifier} manages to recognize a set of $\texttt{cat} (\alpha, S = 0)$ states, with $\mu = 0$ and $\abs{\alpha} \in [2, 3]$, as more and more photons are lost. Before any photons are lost, the softmax probabilities for different labels show that the \texttt{Classifier} assigns the highest probability to the label \texttt{cat}. After $\sim \unit[70]{\%}$ of the photons have been lost, the probability of the state being classified as a $\texttt{cat}$ decreases and the labels $\texttt{coherent}$ and $\texttt{binomial}$ become equally probable. It is an interesting question whether these probabilities reflect the characteristics of the state in such a way that it could be used as a starting point for reconstruction. When almost all the photons are lost, the classification label is always \texttt{coherent}.

Even though we did not include any photon-loss noise during the training phase, the $\texttt{Classifier}$ is still able to identify \texttt{cat} states after many photons have been lost. It should be noted that once photons have been lost, it is not certain that the state can be considered a \texttt{cat} state anymore. A distinctive feature of \texttt{cat} states is the interference between the coherent states making up the superposition state. This interference results in zero probability of odd photon numbers in the state [see the left panel in ~\figpanel{fig:photon-loss}{c}]. Once photon loss starts acting on the state, these occupation probabilities become non-zero [see the middle panel in \figpanel{fig:photon-loss}{c}], but the \texttt{Classifier} network can still identify general features leading it to classify the data with the label \texttt{cat}. However, once more photons have been lost, the state ceases to be a $\texttt{cat}$ state and is classified as a $\texttt{coherent}$ state.

We note that the results presented here for classification under photon loss may be different if the network instead is trained on data in the form of Wigner functions (see \secref{sec:measurements}) instead of Husimi $Q$ functions. The Wigner function for \texttt{cat} states has characteristic interference fringes, some with negative values, between the coherent-state blobs. These features are not clearly seen in the Husimi $Q$ function; it only takes very small non-zero values ($\sim 10^{-4}$) between the two coherent blobs in a $\texttt{cat}(\alpha = 2, S = 0, \mu = 0)$ state. Another approach to identify the lossy \texttt{cat} states better would be to train a classifier to distinguish \texttt{cat} states and mixtures of coherent states from the Husimi $Q$ function. Just like the \texttt{Classifier}, this does not require explicitly specifying criteria for what is or is not a \texttt{cat}, but works in the spirit of ``Software 2.0"~\cite{Karpathy2017} --- replacing explicit programming with learning from data. 


\subsubsection{Classification in the presence of additive Gaussian noise}
\label{subsub:results-gaussian-noise}

\begin{figure}[]
\centering
\includegraphics[width=0.98\linewidth]{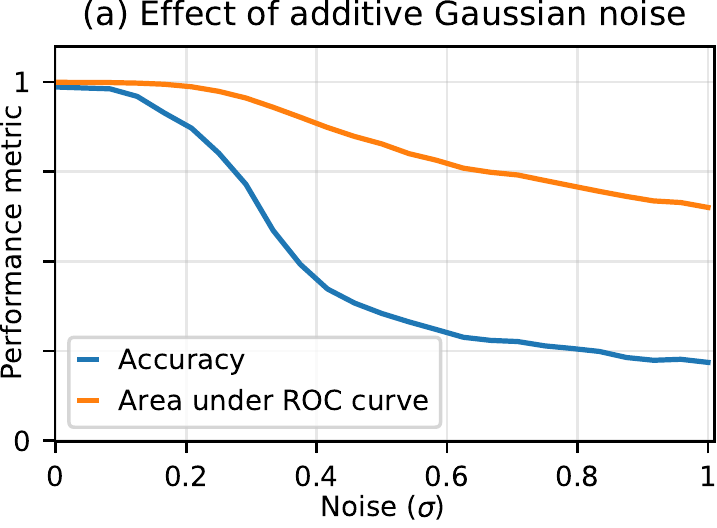}
\vspace{2mm}

\includegraphics[width=0.98\linewidth]{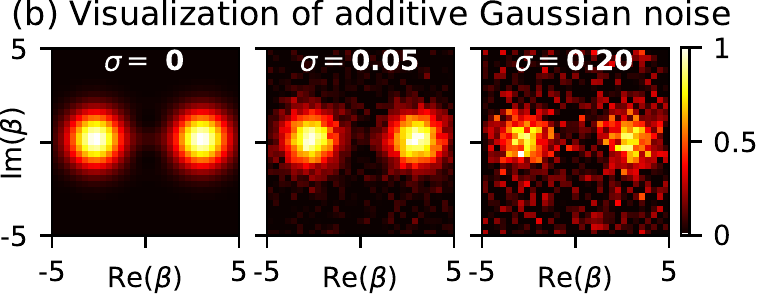}
\vspace{2mm}
\caption{$\texttt{Classifier}$ performance in the presence of additive Gaussian noise.
(a) Accuracy (blue) and area under the ROC curve (orange) as a function of the level of additive Gaussian noise. The noise is added to the dataset used in \figref{fig:confusion}.
(b) Effect of additive Gaussian noise on a $\texttt{cat}(\alpha = 2, S = 0)$ state. The distortion of the state starts to become significant between $\sigma = 0.05$ and $\sigma = 0.2$, which is when the accuracy of the \texttt{Classifier} starts to drop below \unit[90]{\%}.
\label{fig:gaussian-additive}}
\end{figure}

Next, we test the performance of the \texttt{Classifier} in the presence of additive Gaussian noise. As explained in \secref{subsubsec:additive_gaussian_noise}, this type of noise models uncertainty in the data due to averaging over a limited number of measurements and binning of data. In \figpanel{fig:gaussian-additive}{a}, we plot the classification accuracy as a function of the standard deviation $\sigma$ of the added Gaussian noise (see \secref{subsubsec:additive_gaussian_noise}). The dataset is the same as that in \figref{fig:confusion}, but with the Gaussian noise added. In \figpanel{fig:gaussian-additive}{b}, we show an example of how the Gaussian noise impacts a \texttt{cat} state in the dataset.

The accuracy of the predictions from the \texttt{Classifier} remains high until $\sigma \approx 0.05$ and then decreases gradually. However, even at $\sigma = 1$, the accuracy is almost \unit[25]{\%}, clearly better than $\sim 1/7$, which is what one would obtain for a random guess among the seven classes. At these high levels of noise, the network can still correctly classify up to $\sim \unit[60]{\%}$ of the $\texttt{fock}$ states, $\sim \unit[30]{\%}$ of the \texttt{coherent} states, and $\sim \unit[55]{\%}$ of the $\texttt{cat}$ states in the test set. However, at such a high level of noise, the $\texttt{Classifier}$ almost always predicts the label as one of $\texttt{fock}$, $\texttt{coherent}$, or $\texttt{cat}$. Hence accuracy is not the best indicator of performance in all scenarios.

Therefore, in addition to the accuracy, we also quantify the \texttt{Classifier} performance by considering the receiver-operating-characteristic (ROC) curve~\cite{Fawcett2006, Brown2006}. The ROC curve for a binary classification problem is a plot of the true positive rate (TPR, the ratio between correctly classified positive labels and the number of real positive labels) versus the false positive rate (FPR, the ratio between false predictions of positive labels and true predictions of negative labels). The area under the ROC curve gives an indication of the discriminative power of the classifier: the area is $1$ for perfect classification and $0.5$ for random guesses. For our multi-class problem, we use the one-vs-rest strategy in Scikit-learn~\cite{Pedregosa2011} to calculate the area under the ROC curve. The result, averaged over all classes, is shown in \figpanel{fig:gaussian-additive}{a}. The area under the ROC curve shows a behaviour similar to the accuracy, but indicates a somewhat better performance than the latter does.


\subsubsection{What does the network see?}
\label{sec:GradCAM}

The \texttt{Classifier} network is a highly nonlinear function that maps data to a label. In order to find the patterns in the data that the network uses to determine the label, we apply gradient-weighted class activation mapping (Grad-CAM)~\cite{Selvaraju2020}. The Grad-CAM method works by fixing a class label and finding the gradients of the score for this target class (before the softmax activation) with respect to the last convolution layer of the network. This is a form of back-propagation that allows us to construct a heatmap of numbers showing which pixels of the input image influence the output the most.

\begin{figure}[t]
\centering
\includegraphics[width=\linewidth]{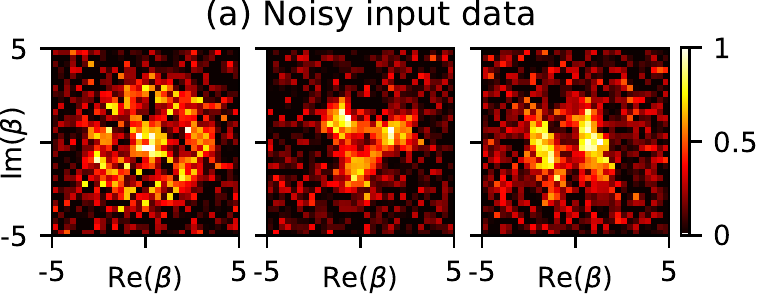}
\vspace{0mm}

\includegraphics[width=\linewidth]{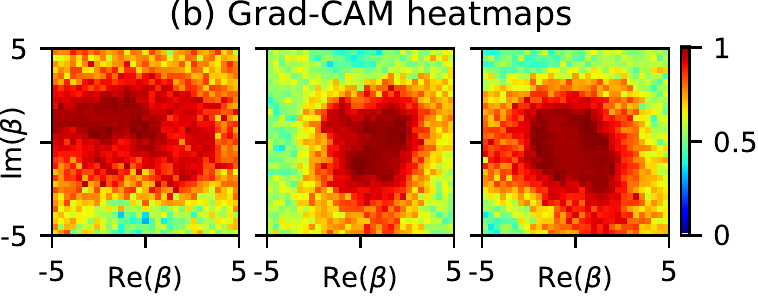}

\vspace{3mm}
\includegraphics[width=\linewidth]{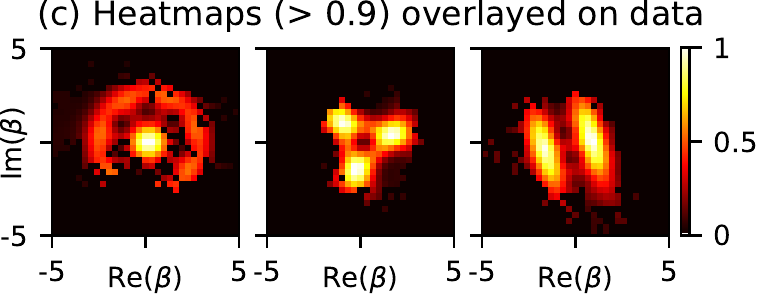}
\caption{Using Grad-CAM to highlight the regions of the input data that the \texttt{Classifier} considers most important for predicting a label.
(a) Input data for three states (from left to right: \texttt{binomial}, \texttt{num}, and \texttt{gkp}) with affine transformations and a constant additive Gaussian noise ($\sigma = 0.2$) for all values of $\beta$ after normalizing the data to the interval $[0, 1]$.
(b) Heatmaps, normalized to the interval $[0, 1]$, constructed with Grad-CAM from the data in (a), showing which parts of the data the \texttt{Classifier} focusses on.
(c) The areas of the data (without the additive Gaussian noise) that appear in the focus when we only show the regions for which the Grad-CAM signals exceed 0.9.
\label{fig:gradcam}}
\end{figure}

In \figpanel{fig:gradcam}{a}, we show three examples of noisy input data, from which we calculate Grad-CAM heatmaps, shown in  \figpanel{fig:gradcam}{b}. These heatmaps are then used in \figpanel{fig:gradcam}{b} to show the parts of the noise-free input data that contribute the most to the classification. Affine transformations (see \secref{subsubsec:affine_noise}) and additive Gaussian noise (see \secref{subsubsec:additive_gaussian_noise}) with $\sigma = 0.2$ have been applied to the input data to simulate an experiment with SPAM errors and little averaging. We chose to only show the parts of the data where the heatmap has high values (exceeding 0.9), to demonstrate that, even in the presence of significant noise, the \texttt{Classifier} makes its decision based on the data in the regions that contain the important patterns characterizing the state. A non-machine-learning way to achieve similar results would be to hand-craft an algorithm that can clean noisy data and detect the regions with a high signal using some boundary-finding algorithm. However, instead of hand-crafting solutions for each type of state and noise, our trained \texttt{Classifier} can easily adapt to a variety of different scenarios.

The Grad-CAM results suggest an interesting possibility for adaptive tomography: using Grad-CAM during the data collection in an experiment to identify regions that are important and then sample from these regions more that from other places. In this way, our \texttt{Classifier} network can identify specific POVMs (defined by displacements $\beta$) that give the most useful data for discriminating optical quantum states.


\subsection{Reconstruction}
\label{subsec:results-reconstruction}


\subsubsection{Impact of loss metric}
\label{sec:loss-functions}

\begin{figure*}[]
\centering
\includegraphics[width=0.8\linewidth]{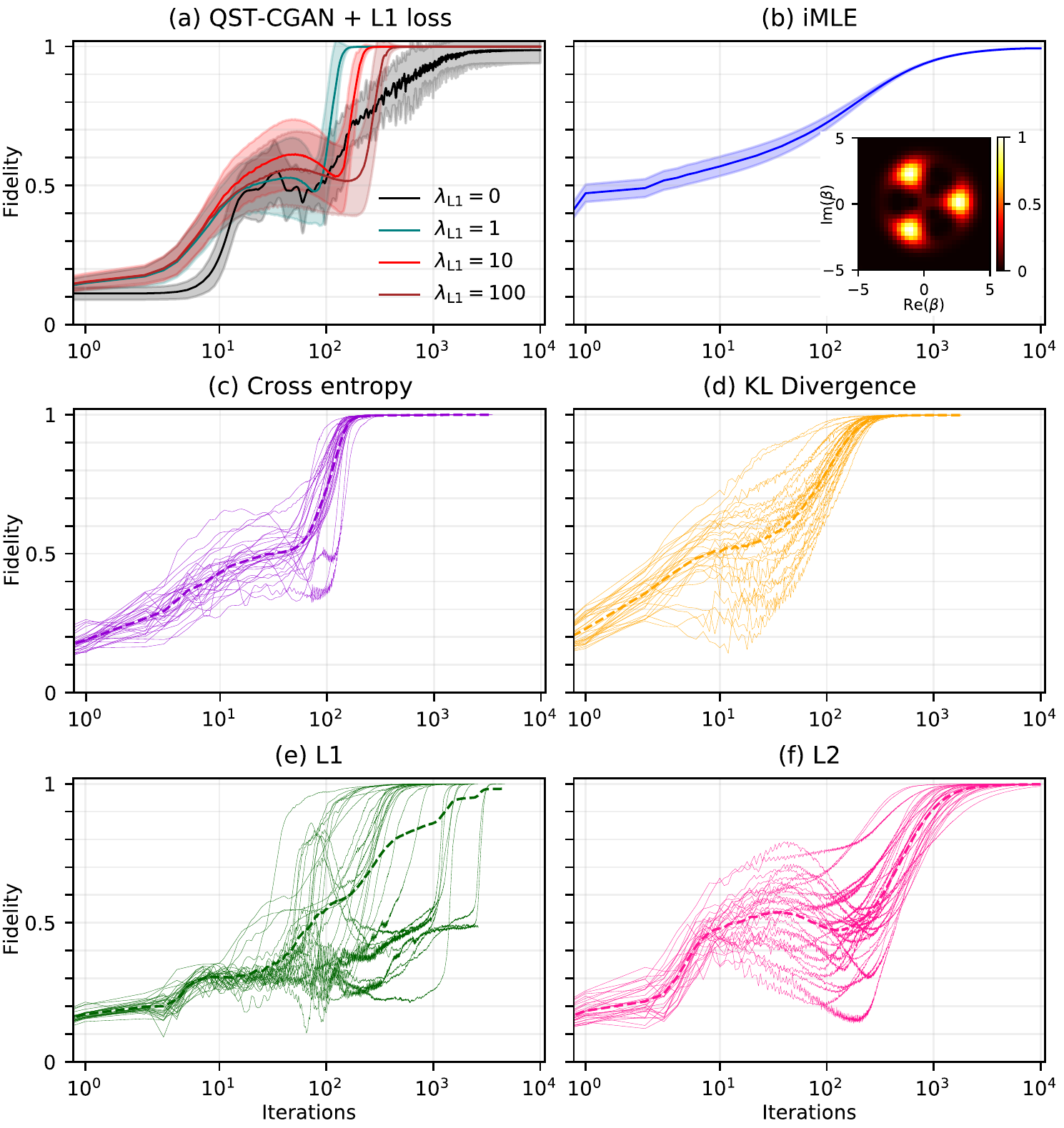}
\caption{The effect of the loss function on reconstruction of $\rho$ using the $\texttt{Generator}$. We use the same data from the Husimi $Q$ function of a $\texttt{binomial}(S = 2, N = 4, \mu = 0)$ state [inset in panel (b)] and repeat the reconstruction with random initializations of the network weights and starting estimate of $\rho$ for iMLE. In each iteration, the weights of the $\texttt{Generator}$ or $\texttt{Discriminator}$ networks are updated using a single step of the Adam optimizer. The learning-rate schedule and optimization hyperparameters are set to the same values (see \secref{sec:reconstruction-training}) for all loss functions in order to achieve a fair comparison. However, further tuning of the parameters for each type of loss function could possibly give better results.
(a) The reconstruction fidelity as a function of iterations for QST-CGAN with various weights of the L1 loss set by the $\lambda_{\rm L1}$ parameter [see \eqref{eq:DiscriminatorL1}]. In each of a total of 30 runs, the weights of the $\texttt{Generator}$ and $\texttt{Discriminator}$ are randomly initialized. The solid lines show the mean and the shaded regions shows one standard deviation from the mean.
(b) The performance of iMLE on the same data. We repeat the reconstruction 30 times and show the mean fidelity (solid lines) and one standard deviation from the mean (shaded region).
(c, d, e, f) Reconstruction fidelities using standard loss functions for the $\texttt{Generator}$: cross-entropy [see \eqref{eq:cross-entropy}], KL divergence, L1, and L2 [see \eqref{eq:L2}]. We show all 30 runs for each loss function with the dashed line showing the mean.
\label{fig:loss-function-binomial}}
\end{figure*}

We first investigate how the choice of loss function affects the performance of our neural-network reconstruction method. In \figref{fig:loss-function-binomial}, we compare the impact of different loss metrics used to train the $\texttt{Generator}$ network. For each loss function, we train the network using the same data. We show the reconstruction fidelity for data from a \texttt{binomial} state reconstructed with six different methods (see \secref{sec:reconstruction-training}). In \figpanel{fig:loss-function-binomial}{a}, we show results for the QST-CGAN with various weights $\lambda_{\rm L1}$ of the L1 loss term in \eqref{eq:DiscriminatorL1}. For all values of $\lambda_{\rm L1}$, including $\lambda_{\rm L1} = 0$, corresponding to pure $\texttt{Discriminator}$ loss, the reconstruction fidelity converges to unity. The convergence is faster with L1 loss added than without it, but a large weight on the L1 part of the loss function leads to worse performance than a moderate weight. The best performance is seen for $\lambda_{\rm L1} = 1$, when the network converges to the correct reconstruction in a little more than 100 iterations, i.e., 100 updates of our estimate for the density matrix. The standard iMLE method, shown in \figpanel{fig:loss-function-binomial}{b}, also converges to unit fidelity, but does so two orders of magnitude slower than the best QST-CGAN in terms of the number of iterations required.

In \figpanels{fig:loss-function-binomial}{c}{f}, we plot the results of training the $\texttt{Generator}$ using the cross-entropy, KL-divergence, L1, and L2 loss functions, respectively. In all cases, the reconstruction fidelity converges to unity. The $\texttt{Generator}$ trained with cross-entropy loss [\figpanel{fig:loss-function-binomial}{c}] displays the fastest convergence, on par with the best QST-CGAN. Training with KL-divergence loss [\figpanel{fig:loss-function-binomial}{d}] gives almost as good results. The L1 [\figpanel{fig:loss-function-binomial}{e}] and L2 [\figpanel{fig:loss-function-binomial}{f}] loss functions result in slower convergence, but still perform better than iMLE for the example considered here. We note that the L1 and L2 loss functions lead to a wider distribution of the number of iterations required for convergence for the same data than any of the other methods.

\begin{figure*}[]
\centering
\includegraphics[width=0.8\linewidth]{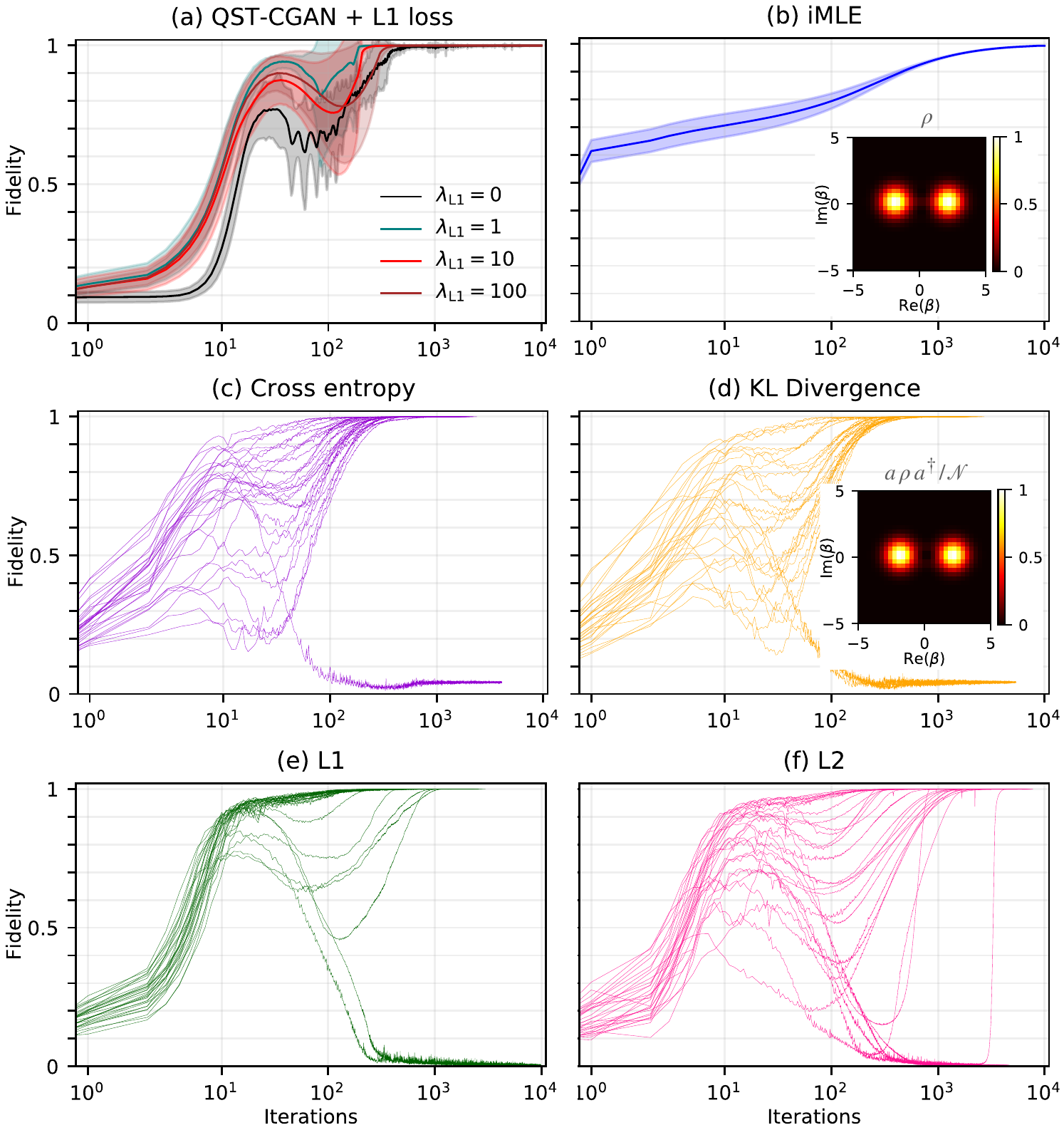}
\caption{The effect of the loss function on reconstruction of a $\texttt{cat}(\alpha=2, S=0, \mu = 0)$ state from Husimi-$Q$-function data [inset in panel (b)]. All hyperparameters, number of runs, and meanings of lines solid lines and shaded regions in the plots are the same as for \figref{fig:loss-function-binomial}.
(a) Performance of the QST-CGAN with various weights of the L1 loss.
(b) Reconstruction fidelities for iMLE on the same data.
(c, d, e, f) Reconstruction performance with standard loss functions. The inset in (d) shows the Husimi $Q$ function of a $\texttt{cat}$ state orthogonal to the one in the inset in (b), which was used to produce the data. The state in (d) is constructed by applying the photon annihilation operator $a$ to the original state in (b).
\label{fig:loss-function-cat}}
\end{figure*}

To ensure that the results in \figref{fig:loss-function-binomial} were not particular to the state used as input data there, we also show the results of reconstruction of a $\texttt{cat}$ state in \figref{fig:loss-function-cat}. The results in \figpanel{fig:loss-function-cat}{a} are similar to those in \figpanel{fig:loss-function-binomial}{a}: the QST-CGAN always converges to unit fidelity, and it does so the fastest when L1 loss is added to the $\texttt{Discriminator}$ loss with weight $\lambda_{\rm L1} = 1$. The main difference to \figpanel{fig:loss-function-binomial}{a} is that the convergence with pure $\texttt{Discriminator}$ loss is considerably faster in \figpanel{fig:loss-function-cat}{a} and is almost as fast as when L1 loss is added. Just as in \figref{fig:loss-function-binomial}, the iMLE method, shown in \figpanel{fig:loss-function-cat}{b}, converges to unit fidelity about two orders of magnitude slower than the best QST-CGAN.

The plots in \figpanels{fig:loss-function-cat}{c}{f} show the results of training the $\texttt{Generator}$ using the cross-entropy, KL-divergence, L1, and L2 loss functions, respectively. Whereas these methods all eventually lead to unit fidelity for the reconstruction in \figref{fig:loss-function-binomial}, here they all sometimes fail and end up in a state giving reconstruction fidelity zero instead. In the cases where they do end up at unit fidelity, the convergence is approximately as fast as in \figref{fig:loss-function-binomial}, perhaps somewhat faster for the L2 loss in \figpanel{fig:loss-function-cat}{f}.

In the cases where the standard loss functions lead the $\texttt{Generator}$ to reconstruct a state with fidelity zero, the reconstructed state is a $\texttt{cat}$ state, shown in the inset of \figpanel{fig:loss-function-cat}{d}, orthogonal to the $\texttt{cat}$ state, shown in the inset of \figpanel{fig:loss-function-cat}{b}, that provides the data. The two $\texttt{cat}$ states have the same $\alpha$ and virtually indistinguishable Husimi $Q$ functions. The only difference between the two is that the correct state has non-zero values in a narrow line along $\text{Im}(\beta) = 0$ between the two prominent lobes in the Husimi $Q$ function, while the orthogonal state has non-zero values at two narrow lines along $\text{Im}(\beta) \approx \pm 0.5$ instead. The differences between the two states are more clearly seen if one plots their Wigner functions instead. We consider reconstruction from Wigner-function samples in \secref{sec:mixed-states-results} and from experimental data in Ref.~\cite{Ahmed2020}.

For the KL divergence in \figpanel{fig:loss-function-cat}{d} and the L1 loss in \figpanel{fig:loss-function-cat}{e}, the $\texttt{Generator}$ network seems to start moving towards one of the two $\texttt{cat}$ states and then eventually converge to that state. However, for the L2 loss in \figpanel{fig:loss-function-cat}{e}, there are some runs where the $\texttt{Generator}$ network reconstructs an orthogonal state (with very low fidelity to the target), but then corrects and jumps to the correct state within a few iterations. In the specific case of a \texttt{cat} state, the orthogonal state is reached by applying the photon annihilation operator $a$ to the correct state. It remains to be explored if the $\texttt{Generator}$ network learns to represent quantum states in a way that it can apply such non-trivial quantum operations to find the correct state from an initially incorrect prediction.

In our attempts to tune the hyperparameters of the training, we have noticed that higher values of the parameters $m_1$ and $m_2$ for the Adam optimizer removes the behaviour seen in \figpanels{fig:loss-function-cat}{c}{f}. Instead, for these values the $\texttt{Generator}$ always finds the correct state and not its orthogonal counterpart, similar to how the QST-CGAN in \figpanel{fig:loss-function-cat}{a} always converges to the correct state. Another way to achieve this convergence could be sampling more around $\text{Re}(\beta)$ to focus on the data that distinguishes the two cat states. In any case, it is noteworthy that the $\texttt{Generator}$ network could possibly apply non-trivial steps to quickly reconstruct the state while the iMLE converges in small steady steps.

To summarize the results in Figs.~\ref{fig:loss-function-binomial} and \ref{fig:loss-function-cat}, the main finding is that the best QST-CGAN reaches unit fidelity orders of magnitude faster than the standard iMLE method. The QST-CGAN performs best when its loss function is an approximately equal mix of L1 loss and the trained $\texttt{Discriminator}$ loss. Further tuning of hyperparameters leads to even better performance in some cases. Among the standard loss functions, cross-entropy loss and KL divergence lead to somewhat better performance for the $\texttt{Generator}$ network than did L1 and L2 loss. However, as we will see in the following subsections, there are other situations, e.g., when the reconstruction is performed in the presence of noise, where these losses give better performance and where a different value for $\lambda_{L1}$ may be more suitable for the QST-CGAN. The fact that different situations seem to require different loss functions is an important argument in favour of the flexibility of the $\texttt{Discriminator}$ loss, which can adapt to the situation, allowing the QST-CGAN to perform well in a more general setting.


\subsubsection{Reconstruction in the presence of additive Gaussian noise}
\label{sec:additive-gaussian-results}

\begin{figure*}[t]
\centering
\includegraphics[width=\linewidth]{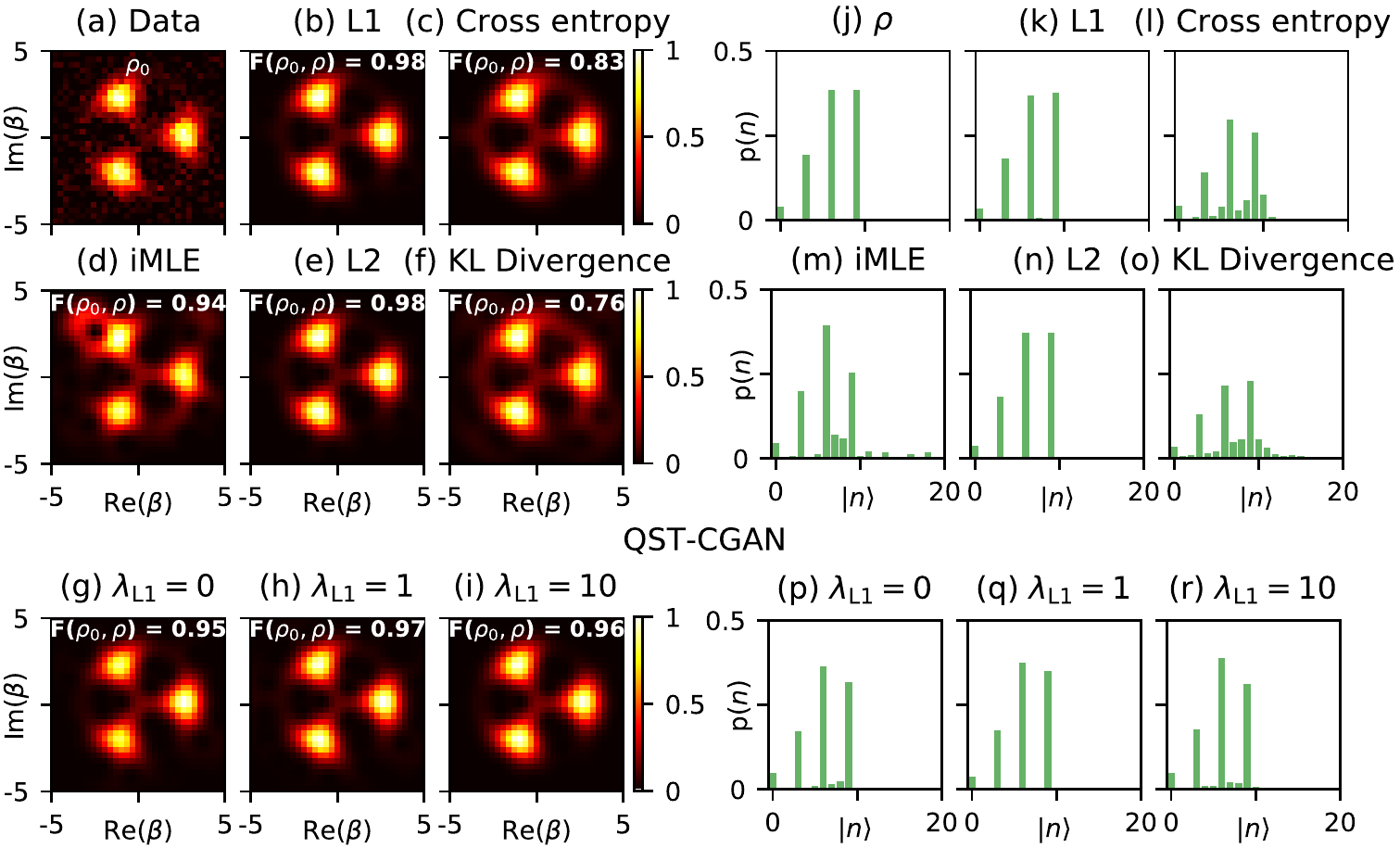}
\caption{Reconstruction of a $\texttt{binomial}(S=2, N=4, \mu = 0)$ state in the presence of additive Gaussian noise.
(a) The Husimi $Q$ function of the state after addition of Gaussian noise at each $\beta$. The random noise is drawn from a standard normal distribution with $\sigma_G = 0.05$ and added after the data has been normalized to the range $[0, 1]$.
(b, c, e, f) Reconstructed Husimi $Q$ functions, without noise added by the \textit{GaussianNoise} layer, using standard loss functions for the $\texttt{Generator}$:  L1, cross-entropy, L2, and KL divergence, respectively.
(d) Reconstructed Husimi $Q$ function using iMLE.
(g, h, i) Reconstructed Husimi $Q$ functions using our QST-CGAN with three different weights of the L1 loss set by $\lambda_{\rm L1}$.
(j) Photon-number occupation probabilities for the data without noise added. 
(k, l, m, n, o, p, q, r) Photon-number occupation probabilities extracted from the reconstructed density matrices corresponding to the Husimi $Q$ functions in (b, c, d, e, f, g, h, i), respectively.
In all reconstructions using neural networks, the hyperparameters for learning were kept the same. For each method, including iMLE, the calculations were stopped after 10,000 iterations.
\label{fig:additive-gaussian-noise-reconstruction}}
\end{figure*}

We now compare how different loss functions affect the neural-network performance in the presence of additive Gaussian noise (see \secref{subsubsec:additive_gaussian_noise}). In \figref{fig:additive-gaussian-noise-reconstruction}, we show representative results of reconstructing a \texttt{binomial} state from Husimi-$Q$-function data where Gaussian noise has been added. For each $\beta$ in the Husimi $Q$ function of the state, we add a random value sampled from a zero-mean Gaussian with standard deviation $\sigma_G = 0.05$. Before adding the noise, we rescale the data to the range $[0, 1]$ by dividing it with the maximum value of the Husimi $Q$ function.

To enable the neural network to learn the state underlying the noisy data, we augment the $\texttt{Generator}$ output with the known noise by introducing a \textit{GaussianNoise} layer. This layer applies the same type of noise to the reconstructed data by sampling from a Gaussian with $\sigma_G = 0.05$ at each gradient-descent step of the Adam optimization. Note that at each step the noise added has the same variance, but differs due to the random sampling. The application of this method in practice requires knowing the type of noise, and its variance, in the experimental setup, but we believe this is feasible to extract.

It might also be possible to simply let the neural network learn the noise. However, applying back-propagation techniques for training requires calculation of gradients with respect to the parameters. The automatic differentiation methods usually employed for gradient calculation in neural networks are not straightforward to apply when such stochastic noise layers are present in the networks. Nevertheless, methods such as the reparameterization trick~\cite{Kingma2014} can still make it possible to learn the noise. However, we have not explored this possibility further in this work.

Looking at the reconstructed Husimi $Q$ functions in \figpanels{fig:additive-gaussian-noise-reconstruction}{b}{i}, it appears that the $\texttt{Generator}$ with L1 or L2 loss and the QST-CGAN with $\lambda_{\rm L1} = \{0, 1, 10\}$ outperform the $\texttt{Generator}$ with cross-entropy or KL divergence loss, and clearly outperform the iMLE. However, a small difference in the appearance of the Husimi $Q$ function does not necessarily mean that two states are similar (compare the orthogonal states depicted in the insets of \figref{fig:loss-function-cat}). We therefore plot, in \figpanels{fig:additive-gaussian-noise-reconstruction}{j}{r}, the photon-number occupation probabilities corresponding to the noiseless data and the reconstructions in \figpanels{fig:additive-gaussian-noise-reconstruction}{b}{i}. The noiseless data has non-zero probabilities for 0, 3, 6, and 9 photons. This is only reproduced well by the $\texttt{Generator}$ with L1 or L2 loss and the QST-CGAN with $\lambda_{\rm L1} = 1$. The QST-CGAN with $\lambda_{\rm L1} = 1$ also reproduces the equal probabilities of 6 and 9 photons in the data better than the QST-CGAN with $\lambda_{\rm L1} = \{0, 10\}$.

\begin{figure*}[t]
\centering
\includegraphics[width=0.8\linewidth]{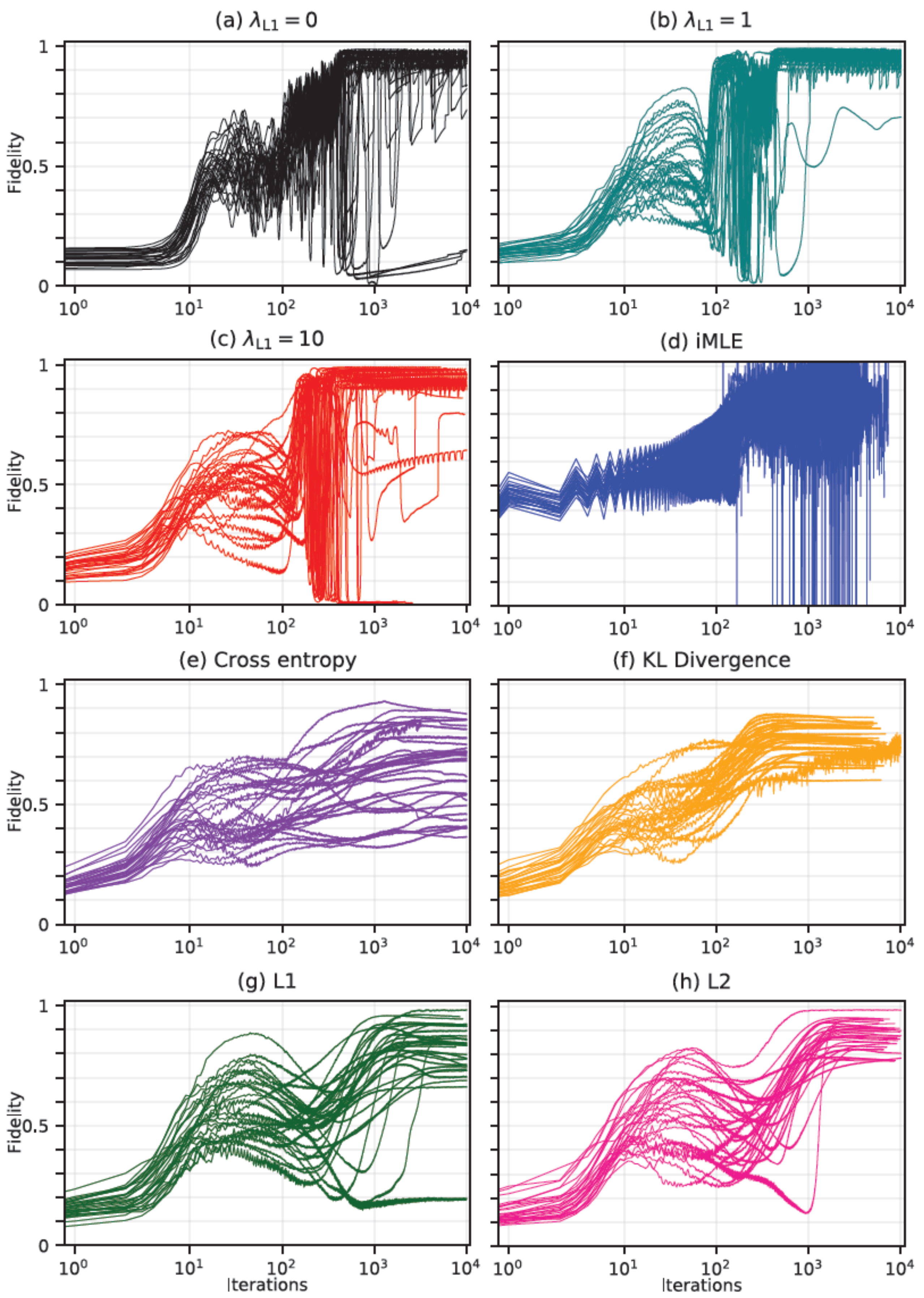}
\caption{The effect of the loss function on reconstruction of the \texttt{binomial} state in \figref{fig:additive-gaussian-noise-reconstruction} with 30 different realizations of the additive Gaussian noise. We show all runs for each loss function. The injected noise has the same variance $\sigma_G = 0.05$ in each run, but is sampled anew for each run.
(a, b, c) Reconstruction fidelity for the QST-CGAN with various weights of the L1 loss.
(d) Reconstruction fidelity for iMLE.
(e, f, g, h) Reconstruction fidelity obtained by training the $\texttt{Generator}$ using standard loss functions: cross-entropy, KL divergence, L1, and L2.
In all the neural-network based reconstructions, we use the same hyperparameters for training in order to have a fair comparison.
\label{fig:gaussian-noise}}
\end{figure*}

To further investigate how different loss functions affect the neural-network performance in the presence of additive Gaussian noise, we plot, in \figref{fig:gaussian-noise}, how the reconstruction fidelity develops, for all reconstruction methods, as a function of the number of iterations for 30 different realizations of the noise in the data (the same \texttt{binomial} state as in \figref{fig:additive-gaussian-noise-reconstruction}). The average reconstruction fidelities and standard deviations are summarized in \tabref{tab:gaussian-fidelities}, where we exclude the reconstructions when the state converges to an orthogonal state [see, e.g., \figpanel{fig:gaussian-noise}{c} for $\lambda_{\mathrm {L1}} = 10$]. Note that we have not tuned the hyperparameters for training, which can lead to improvements for all methods.

\begin{table}[]
\centering
\caption{Mean and standard deviations for fidelities $F$ reached for reconstruction of the \texttt{binomial} state in \figref{fig:additive-gaussian-noise-reconstruction} in the presence of additive Gaussian noise ($\sigma = 0.05$). We consider 30 different sets of noise for each type of loss function. The full trajectories for the fidelities, as each method iteratively updates the estimate of the density matrix, are shown in \figref{fig:gaussian-noise}.
\label{tab:gaussian-fidelities}}
\renewcommand{\arraystretch}{1.25}
\renewcommand{\tabcolsep}{0.15cm}
\begin{tabular}{ | l | c | c |}
\hline
\textbf{Loss} & \textbf{Mean} $F$ & \textbf{Std} ($F$) \\
\hline
QST-CGAN ($\lambda_{\mathrm L1} = 0$) & 0.85 & 0.24\\
\hline
QST-CGAN ($\lambda_{\mathrm L1} = 1$) & 0.95 & 0.05\\
\hline
QST-CGAN ($\lambda_{\mathrm L1} = 10$) & 0.93 & 0.07\\
\hline
Cross-entropy & 0.65 & 0.15\\
\hline
KL-Divergence & 0.76 & 0.06\\
\hline
L1 & 0.81 & 0.14\\
\hline
L2 & 0.87 & 0.05\\
\hline
\end{tabular}
\end{table}

However, it is clear from \figref{fig:gaussian-noise} that the average reconstruction fidelities alone do not give a complete picture of the performance. Looking at the best runs for each method, we see that the QST-CGANs, the iMLE, and the $\texttt{Generator}$ with L1 or L2 loss all are able to reach fidelities very close to 1, while the the $\texttt{Generator}$ with cross-entropy or KL-divergence loss never ends up above fidelity $0.9$. Looking at the spread of results, we see that the iMLE is very unstable, while the QST-CGAN and the $\texttt{Generator}$ with L1 loss have a small number of runs ending up at a very low fidelity. The $\texttt{Generator}$ with L2 loss appears to produce high-fidelity reconstructions with the greatest consistency.

Finally, we can compare how fast the different methods reach high reconstruction fidelity. Here, we see the same trend in \figref{fig:gaussian-noise} as in Figs.~\ref{fig:loss-function-binomial} and \ref{fig:loss-function-cat}: the QST-CGAN is faster than the $\texttt{Generator}$ trained with standard loss functions, and the fastest QST-CGAN is the one with $\lambda_{\rm L1} = 1$.

To summarize the results in Figs.~\ref{fig:additive-gaussian-noise-reconstruction} and \ref{fig:gaussian-noise}, the QST-CGAN and the $\texttt{Generator}$ trained with L1 or L2 loss outshone the other methods when reconstructing a state in the presence of additive Gaussian noise. For the $\texttt{Generator}$ trained with standard loss functions, the results were thus different from the noiseless case in \secref{sec:loss-functions}, when training with cross-entropy or KL-divergence loss gave better results. When comparing the best QST-CGAN, the one trained with $\lambda_{\rm L1} = 1$, to the best $\texttt{Generator}$ trained with standard loss functions, the one trained with L2 loss, we find similar performance in terms of reconstruction fidelities, but the QST-CGAN is faster to reach a good reconstruction.

The errors in reconstruction using the cross-entropy and KL-divergence loss are expected, because these loss functions, similar to iMLE, assume the incorrect likelihood [\eqref{eq:likelihood}] for the data, which does not include the Gaussian error model of \eqref{eq:likelihood-gaussian}. The QST-CGAN reconstruction performs better than these methods since it has the flexibility to learn an appropriate loss function. We only provide the overall objective of making the reconstructed statistics similar to the data.

The reason for the good performance using the L2 loss can be further understood from arguments presented in a recent work on image denoising --- \textit{Noise2Noise}~\cite{Lehtinen2018}. There, the authors note that the expectation value for the loss function when using L2 loss remains unchanged if the targets are replaced by random numbers distributed such that their expectation value matches the target. The crucial insight is that ``the training targets of a neural network can be corrupted with zero-mean noise without changing what the network learns''. Similarly, the L1 loss recovers the median values of the targets and is thus not affected by outliers.

As a final remark, we note that an important factor to consider is that fidelity may not be the best metric to compare results, since sometimes completely random quantum states can have a high fidelity with a desired state~\cite{Montanaro2007}. Moreover, for continuous-variable quantum states such as the optical states considered here, several states with high overlap can have very different characteristics~\cite{Montanaro2007, Mandarino2016}. 


\subsubsection{Reconstruction in the presence of Gaussian convolution noise}
\label{sec:conv-gaussian-results}

\begin{figure*}[ht]
\centering
\includegraphics[width=\columnwidth]{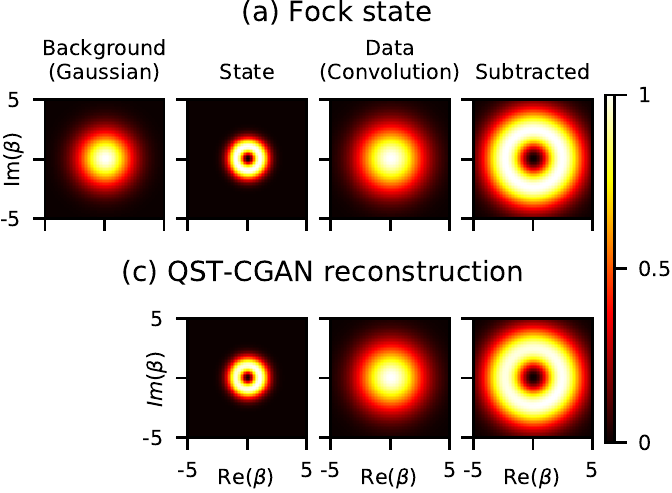}
\includegraphics[width=\columnwidth]{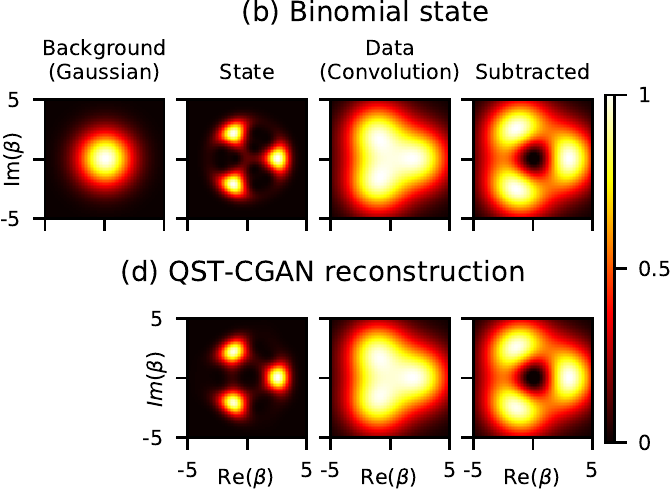}
\caption{Reconstruction in the presence of Gaussian convolution noise. We assume that the noisy amplification channel has a thermal noise with $n_{\text{th}} = 5$ photons  (see \secref{subsubsec:conv_noise}). This noise is convoluted with the data from the Husimi $Q$ function in 
(a) a $\texttt{fock} (n = 1)$ state and 
(b) a $\texttt{binomial} (S=2, N=4)$ state.
We had to use an $81 \times 81$ grid for the data since we noticed that the convolution operation leads to effects such as loss of radial symmetry in (a) when using a grid of $32 \times 32$. The image obtained by subtracting the background from the convoluted data shows the symmetries of the state. 
(c, d) Reconstruction of the underlying states in (a, b) by the QST-CGAN ($\lambda_{\rm L1} = 10$) where the $\texttt{Generator}$ output is convoluted with the same background noise. We recover the underlying state from the \textit{DensityMatrix} layer of the $\texttt{Generator}$. Note that even though the convoluted outputs and subtracted counts match the data well in both cases, the fidelity between the underlying reconstructed state itself and the true underlying state is $1$ in (c), but only $0.45$ in (d).}
\label{fig:gaussian-convolution}
\end{figure*}

The additive Gaussian noise discussed in the preceding section models statistical errors due to a low signal-to-noise ratio. Such noise can be reduced (averaged out) by taking more measurements. However, in many cases we have other types of noise that can corrupt the data. Removing such noise in the context of image processing constitutes an inverse problem that is often difficult or ill-posed and requires regularization techniques~\cite{Karl2005}.

We now show how to deal with one such type of noise: Gaussian convolution noise (see \secref{subsubsec:conv_noise}) due to a linear amplification channel~\cite{Eichler2012}. In such a setup, a background noise, which usually is easy to estimate, corrupts our signal via a convolution operation. Similar to \secref{sec:additive-gaussian-results}, we consider this known background noise as an input to the $\texttt{Generator}$ and augment the $\texttt{Generator}$ with a \textit{GaussianConv} layer such that the $\texttt{Generator}$ output is convoluted with the noise in the same way as the data. This noise layer is not learned, but fixed to the pre-determined background noise. The addition of the noise layer forces the $\texttt{Generator}$ network to learn a density matrix $\rho'$ that can generate similar statistics as the data after convolution with the background noise.

In \figpanel{fig:gaussian-convolution}{a}, we show the results of reconstructing a single-photon Fock state from the Husimi-$Q$-function data after convolution with a background noise arising due to the amplification channel being in a thermal state. In the simulations considered in preceding sections, we used a $32 \times 32$ grid of measurements. However, this coarse grid led to numerical aberrations in the convolution operation. For the present section, we therefore considered an $81 \times 81$ grid instead. The results show that the underlying single-photon state is reconstructed perfectly with unit fidelity by a QST-CGAN with $\lambda_{\rm L1} = 10$ despite the presence of significant noise .

However, since the inverse problem can be ill-posed, it is also possible to obtain a result that reconstructs the data well without getting the underlying state right. In \figpanel{fig:gaussian-convolution}{b}, we show the one such reconstruction using a \texttt{binomial} state in the presence of the same Gaussian convolution noise as in \figpanel{fig:gaussian-convolution}{a}. The reconstructed density matrix gives rise to measurement statistics that match the measured (simulated) data exactly. However, the state itself is incorrect with a fidelity of just $0.45$. We note that the symmetries of the state are captured in the reconstruction, but due to the convolution operation, the information of the exact state is lost and the inversion is not unique.


\subsubsection{Reconstruction of mixed states}
\label{sec:mixed-states-results}

So far, we have only considered pure quantum states, where the density matrix has rank $r = 1$. However, in real experiments, we will almost always be dealing with mixed states. Such states may be harder for a neural network to handle, since they do not admit as compact a representation as a pure state, which can be written $\rho = \ketbra{\psi}{\psi}$. In this section, we therefore discuss how the QST-CGAN method performs for mixed states with rank $r > 1$. 

\begin{figure*}[ht]
\centering
\includegraphics[width=0.95\linewidth]{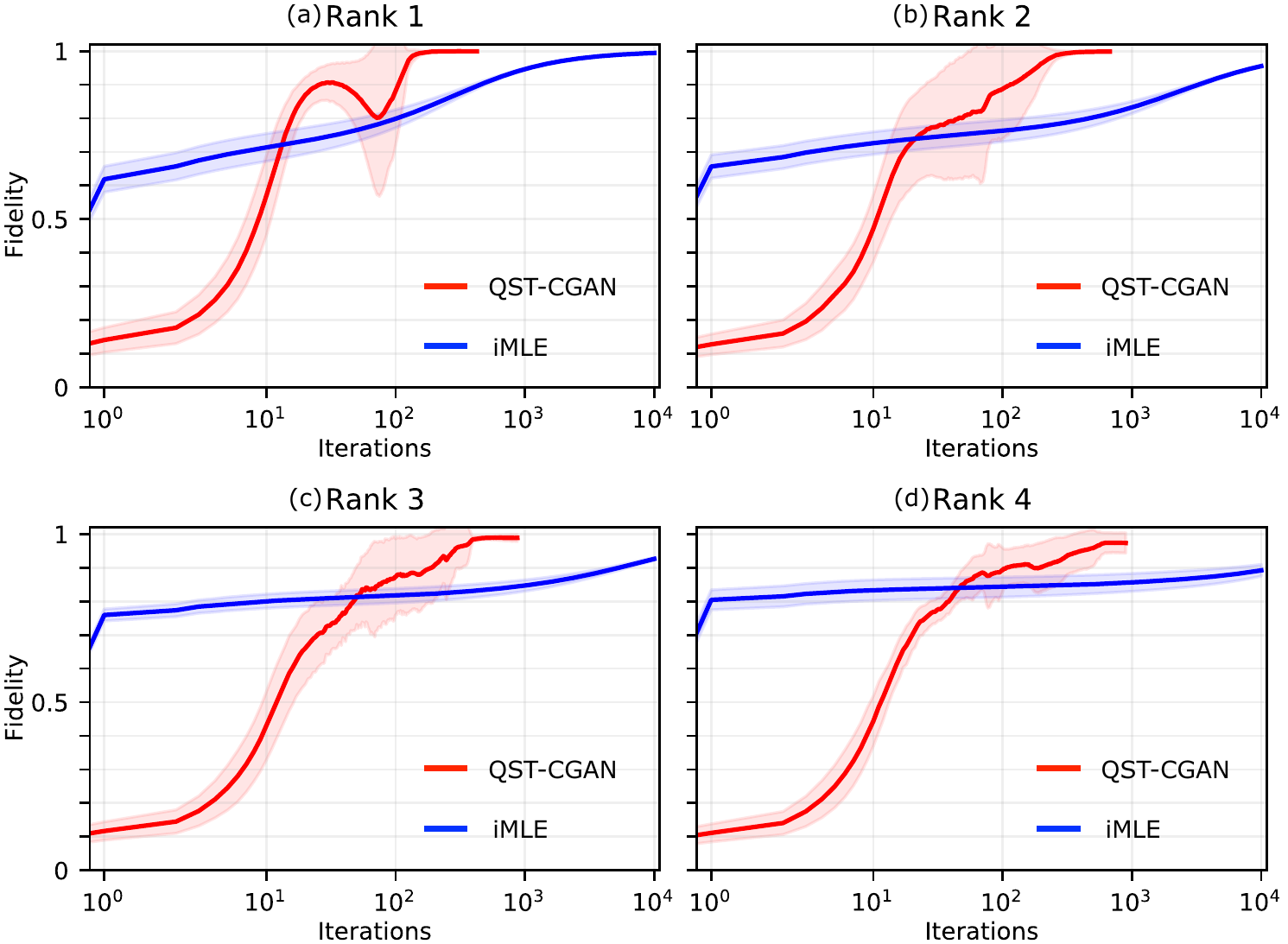}
\caption{Reconstruction of a mixture $\rho = 0.8 \texttt{cat} + [0.2/(r-1)] \sum_{n=0}^{r-2}\texttt{fock}(n)$ of $\texttt{cat}(\alpha = 2, S = 0, \mu = 0)$ (the state used in \figref{fig:loss-function-cat}) and $\texttt{fock}(n)$ states, where $r \ge 2$ denotes the rank. For $r = 1$, the state is just $\texttt{cat}(\alpha = 2, S = 0, \mu = 0)$. The input data is the Husimi $Q$ function of $\rho$ measured in a $32 \times 32$ grid. 
(a, b, c, d) States with ranks 1, 2, 3, and 4, respectively. The solid lines show the mean and the shaded regions show one standard deviation from the mean for the QST-CGAN (red) and iMLE (blue) over 15 reconstructions. In each repetition, we use the same data, but start from a different random initial state for iMLE and random weights for the QST-CGAN. We choose the weight $\lambda_{\rm L1} = 1$ for the QST-CGAN and keep all other training hyperparameters the same as used for previous results and described in \secref{sec:reconstruction-training}. The QST-CGAN runs are stopped when they have converged on a reconstructed state.
\label{fig:mixed-states}}
\end{figure*}

In a realistic experiment, it is reasonable to assume that the mixed state will have a dominant part, e.g., a target state which decoheres due to photon loss. Figure~\ref{fig:mixed-states} shows results for the QST-CGAN reconstruction on a mixed state with a $\texttt{cat}$ state (the same state as in \figref{fig:loss-function-cat}) being the dominant component. The figure shows that the QST-CGAN can reconstruct such a mixed state easily for ranks up to $r = 4$ with close to unit fidelity ($\ge .99$). For ranks $1$ and $2$, the QST-CGAN method converges almost two orders of magnitude faster than iMLE. As the rank increases, both QST-CGAN and iMLE show a slower convergence. Although we did not run the iMLE for enough iterations to be certain, the increase in the number of iterations required for convergence appears to be greater for iMLE than for the QST-CGAN.

\begin{figure}[ht]
\centering
\includegraphics[width=\linewidth]{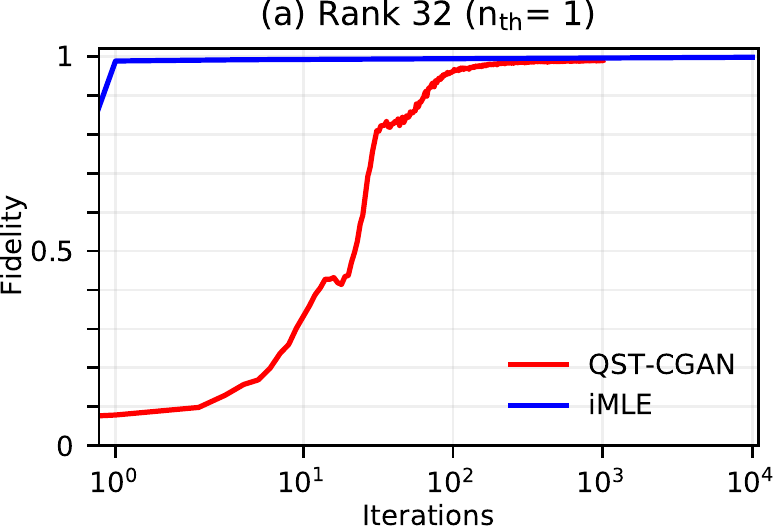}
\vspace{0.1mm}

\includegraphics[width=\linewidth]{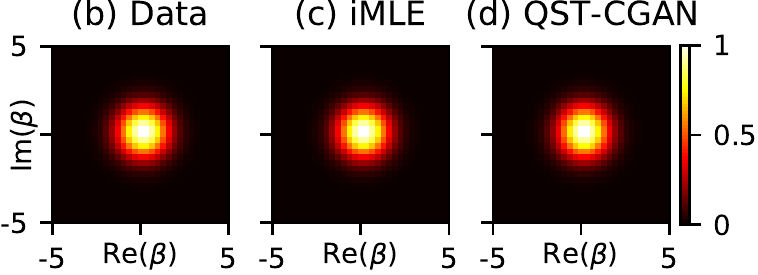}
\vspace{0.1mm}

\includegraphics[width=\linewidth]{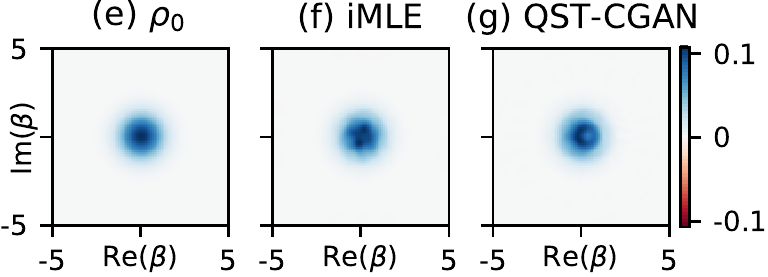}
\caption{Reconstruction of a thermal state with mean photon number $n_{\rm th} = 1$. The QST-CGAN method uses the weight $\lambda_{\rm L1} = 1$; all other training hyperparameters are kept the same as for previous results and described in \secref{sec:reconstruction-training}.
(a) The fidelity of the reconstructed state as a function of the number of iterations for iMLE (blue) and QST-CGAN (red).
(b, c, d) Reconstructed data compared to the data used for obtaining the underlying density matrix. We use the Husimi $Q$ function measured in a $32 \times 32$ grid as the input.
(e, f, g) Wigner function of the underlying thermal state compared to the Wigner functions obtained from the reconstructions. The reconstructed Wigner functions do not match the smooth nature of the Wigner function obtained from the underlying state. 
\label{fig:mixed-states-thermal}}
\end{figure}

\begin{figure*}[]
\centering
\includegraphics[width=0.9\linewidth]{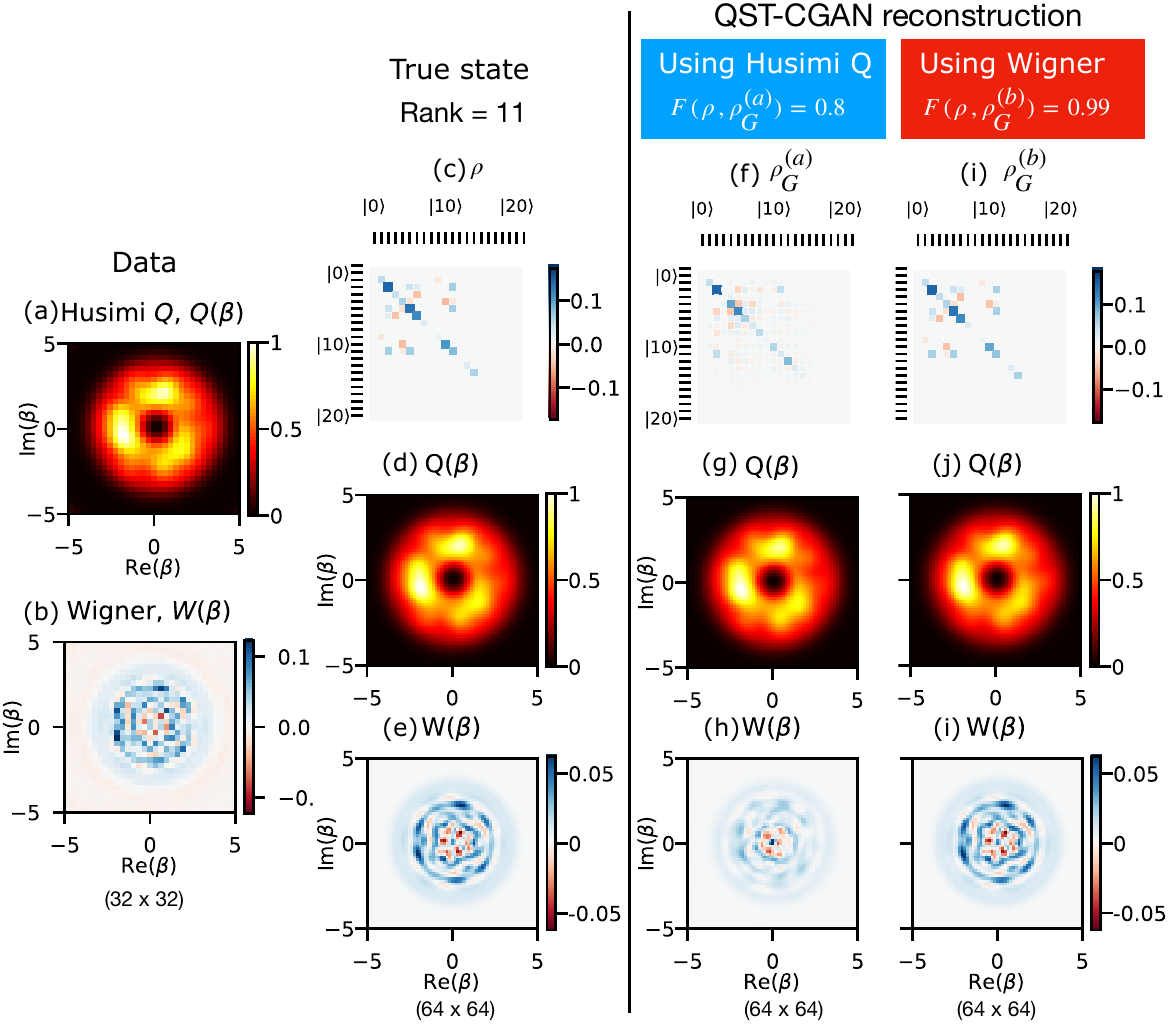}
\caption{Reconstruction with a QST-CGAN of a random mixed state of rank 11 from Husimi $Q$ and Wigner functions sampled on a $32 \times 32$ grid.
(a, b) Data used for the reconstruction: the Husimi $Q$ and Wigner functions.
(c) Hinton plot (see \figref{fig:zoo}) of the underlying density matrix.
(d, e) Husimi $Q$ and Wigner functions for a $64 \times 64$ grid computed from the underlying state to show finer features not present in the data that is fed to the neural network.
(f, g, h) Reconstruction results using the Husimi-$Q$-function data in (a) as input for the QST-CGAN.
(i, j, k) Reconstruction results using the Wigner-function data in (b) as input for the QST-CGAN.
\label{fig:mixed-states-random}}
\end{figure*}

To explore further for higher ranks, we consider in \figref{fig:mixed-states-thermal} the reconstruction of a full-ranked ($r = 32$) thermal state with a mean photon number $n_{\rm th} = 1$. Here, the iMLE method converges very fast, almost instantaneously, while the QST-CGAN requires several hundred iterations. Although both methods reconstruct the state with a high fidelity $\ge 0.99$, the photon-number populations of the reconstructed state do not exactly match the expected super-poissonian distribution for thermal states for the higher photon numbers (the tail of the distribution), neither for iMLE nor for QST-CGAN. The Husimi $Q$ function of the reconstructed states match well in \figpanel{fig:mixed-states-thermal}{b}, but the Wigner functions for the iMLE and QST-CGAN methods do not match the smooth Wigner function for the thermal state in \figpanel{fig:mixed-states-thermal}{c}. However, changing the input data for reconstruction to the Wigner function (displaced parity measurements; see \secref{sec:measurements}) can lead to a better reconstruction as we discuss in \figref{fig:mixed-states-random}. For the QST-CGAN, this change is as simple as replacing the input measurement operators to the $\texttt{Generator}$ from projections on the $\texttt{coherent}$ state (Husimi $Q$) to displaced parity operators (Wigner).

Having explored reconstruction performance for low- and high-rank density matrices, we next turn to intermediate rank. In \figref{fig:mixed-states-random}, we consider a random density matrix of rank $11$ and show its reconstruction from both types of input data (Husimi $Q$ and Wigner). Here, the difference between using the two types of data becomes clear. With the QST-CGAN method, we obtain a reconstruction fidelity of $0.8$ using Husimi $Q$ function and $\sim 0.99$ when using the Wigner function. In \figpanel{fig:mixed-states-random}{h}, we can clearly see that details of the Wigner function for the true state in \figpanel{fig:mixed-states-random}{e} are not captured when we take Husimi $Q$ as our data, even though the reconstructed Husimi $Q$ function in \figpanel{fig:mixed-states-random}{g} matches perfectly with the data in \figpanel{fig:mixed-states-random}{d}. However, there are big differences in how different ML models perform, within the limitations of the data; there is no silver bullet. In this particular case, we can argue that the Husimi $Q$ represents a convolution over the Wigner function~\cite{Andreev2011} and therefore using it for reconstruction could be ill-posed~\cite{Karl2005}.


\subsubsection{Data reduction}
\label{sec:data-reduction-results}

In Ref.~\cite{Ahmed2020}, we show that for a particular pure \texttt{cat} state, the QST-CGAN method requires much fewer data points, $\sim 100$, for reconstruction than the iMLE method, which requires more than 10,000. In Ref.~\cite{Sych2012}, it is argued how homodyne tomography can be IC when the number of independent quadratures measured is equal to the dimension $N$ of the density matrix. More specifically, IC requires $N$ quadratures to be measured, each of which can be discretized into $2 N - 1$ bins. Therefore, a full-rank density matrix of dimension $N$ requires $O (N^2)$ measurements in the phase space for IC. However, for low-rank states the data requirements can scale as $O (rN)$~\cite{Flammia2012}. These arguments suggest that for states described with density matrices of dimension $N = 32$, thousands of measurements are required for IC when considering full-rank states. However, for low-rank or rank-$1$ pure states, the number of data points for IC could be much smaller ($\propto 32 r$). Note that the IC limit does not necessarily specify which measurements are important and give maximum information; the limit also depends on the density-matrix dimensions, which we can set to have different cut-offs for optical quantum states. Our QST-CGAN approach consistently required only $\sim 100$ measurements for reconstruction of pure ($r=1$) states using a random set of measurement settings.

In this section, we benchmark the QST-CGAN performance further by testing how much data it needs to reconstruct states of higher rank. In general, a density matrix of size $N \times N$ with full rank $r = N$ is specified by $N^2 - 1$ real numbers. The number of parameters that needs to be determined during reconstruction is significanly reduced if the state is pure or if we have some prior information about the state~\cite{Sych2012}. For example, if we know that the state that we are reconstructing is a thermal state, then even if the density matrix is full rank ($r = N$), we only need to estimate a single parameter, the mean photon number $n_{\rm th}$, to reconstruct the state. Such priors can thus make it easy to reconstruct the state with data from only a few measurements. Similarly, the analyticity of the Husimi $Q$ function makes it possible to apply other reconstruction methods, e.g., Lagrange interpolation~\cite{Landon-Cardinal2018}, which sample from the Husimi $Q$ function and obtain the exact density matrix without requiring any iterations. Similarly, Ref.~\cite{Nehra2020} reconstructed a state description in the Fock basis using Wigner-function-overlap measurements and semi-definite programming, requiring less data.

\begin{figure*}[]
\centering
\includegraphics[width=0.49\linewidth]{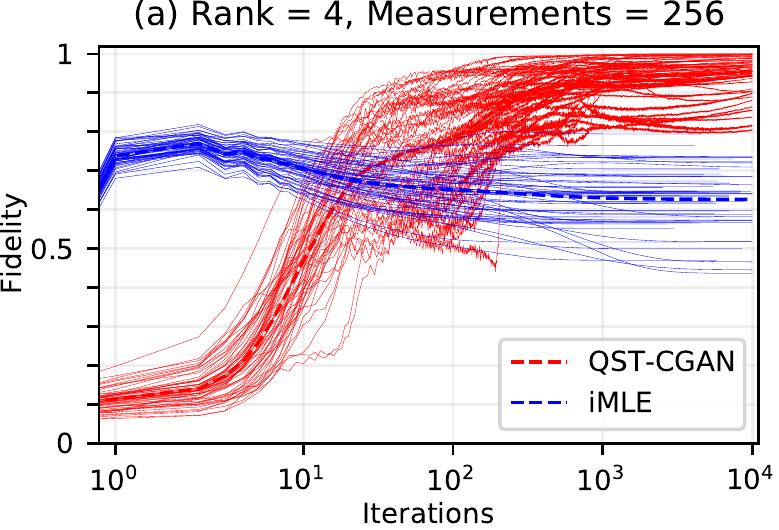}
\includegraphics[width=0.49\linewidth]{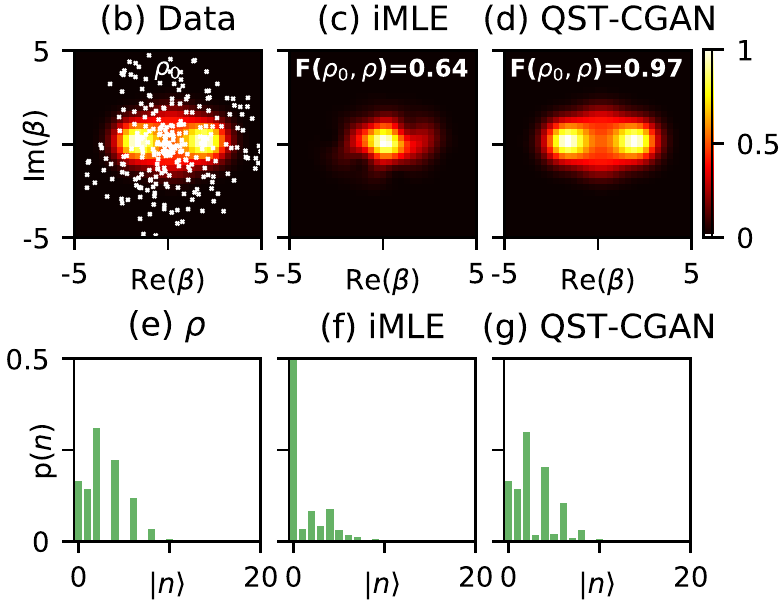}
\caption{Reconstruction of a rank-4 mixture of $\texttt{cat}(\alpha = 2, S = 0, \mu = 0)$ and $\texttt{fock} (n)$ states (the mixture is constructed using the same formula as in \figref{fig:mixed-states}) from input data consisting of 256 points of the Husimi $Q$ function (instead of the full $32 \times 32 =  \text{1,024}$ points used in preceding sections).
(a) Reconstruction fidelity of the QST-CGAN (with $\lambda_{\rm L1} = 1$; red) and iMLE (blue) where we plot all trajectories for $36$ different reconstructions and show the mean with dashed lines. In each reconstruction, we randomly selected a set of $\beta$ values from the phase space and reconstruct the state where the QST-CGAN weights and the initial state for iMLE are reinitialized randomly. 
(b, c, d) Comparison between the data and the reconstructed Husimi $Q$ function given by iMLE and QST-CGAN for one selection of input data points (white points in the left panel).
(e, f, g) Photon-number occupation probabilities from the (reconstructed) density matrices in (b).
\label{fig:data-reduction-example}}
\end{figure*}

In \figref{fig:data-reduction-example}, we show how reconstructing a rank-4 state from a random selection of $256$ points of the Husimi $Q$ function fails for iMLE, which gets stuck. The convergence of iMLE is not guaranteed since there could be steps which strictly reduce the likelihood, producing cycles where the method does not improve its estimate of the density matrix~\cite{Rehacek2007}. The QST-CGAN, on the other hand, reconstructs the state almost perfectly, as shown in \figpanel{fig:data-reduction-example}{d}. However, the QST-CGAN requires more than 1,000 iterations to converge, which is more than what was needed when it reconstructed the same state using all data [see \figpanel{fig:mixed-states}{d}].

\begin{figure}[ht]
\centering
\includegraphics[width=0.88\linewidth]{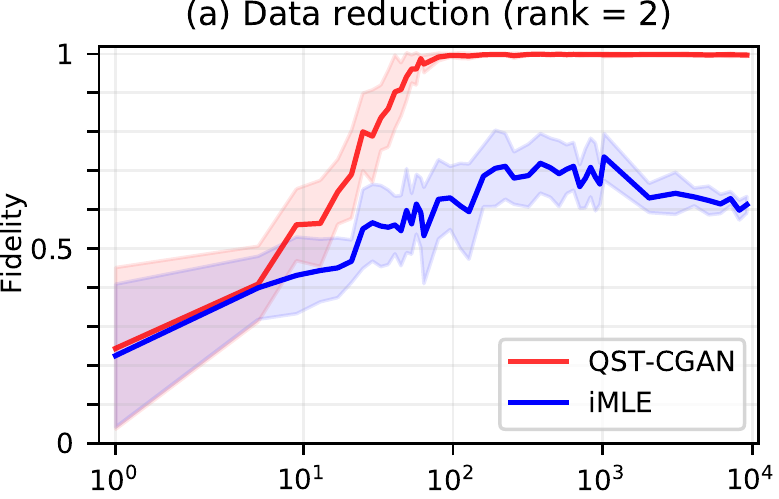}
\vspace{3mm}

\includegraphics[width=0.88\linewidth]{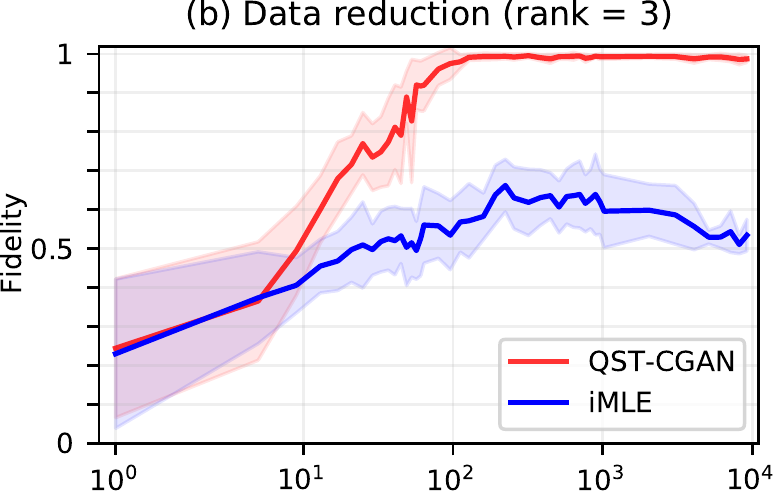}
\vspace{3mm}

\includegraphics[width=0.88\linewidth]{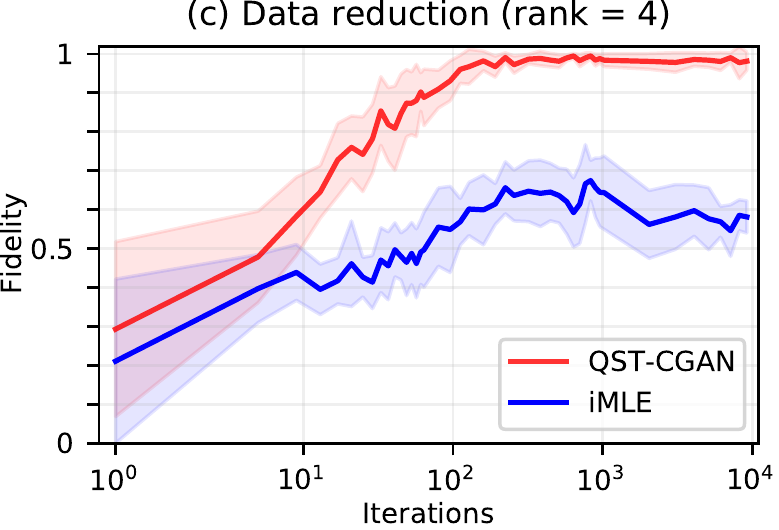}
\vspace{3mm}

\caption{Reconstruction of a mixture of $\texttt{cat}(\alpha = 2, S = 0, \mu = 0)$ and $\texttt{fock}(n)$ states (the mixture is constructed using the same formula as in \figref{fig:mixed-states}) with a reduced number of measurements.
(a, b, c) Reconstruction fidelities for ranks 2, 3, and 4, respectively, for QST-CGAN (red) and iMLE (blue). The solid lines show the mean and the shaded regions show one standard deviation from the mean. The fidelity shown is the one reached after a certain convergence criteria set by a tolerance value. We choose a tolerance such that if the average fidelity in $100$ iterations does not change by $10^{-5}$ over $5$ steps (i.e., $500$ iterations) we stop the reconstruction.
\label{fig:data-reduction}}
\end{figure}

In \figref{fig:data-reduction}, we show the reconstruction fidelity with QST-CGAN and iMLE for mixed states of rank $r = 2, 3, 4$ as the number of measurements (data points; values of the Husimi $Q$ function at different $\beta$) is reduced. We choose the $\beta$ values at random inside the circle $\abs{\beta} = 5$. We note that there could be better ways to choose points to sample the Husimi $Q$ function, e.g., the so-called Padua points discussed in Ref.~\cite{Landon-Cardinal2018}.

The QST-CGAN clearly outperforms the iMLE in terms of the amount of data needed for reconstruction in \figref{fig:data-reduction}. The QST-CGAN reaches fidelity close to unity with somewhat less than 100 data points, around 100 data points, and a little more than 100 data points, for states of rank 2, 3, and 4, respectively.

This slow growth in the number of data points needed appears consistent with previous results showing that reconstruction of low-rank states in the best case can be done with $\propto r N$ data points~\cite{Kalev2015}. Meanwhile, the iMLE cannot reconstruct $\rho$ even when given a large number of data points. However, this is not just due to the lack of information but also due to the random selection of the data points themselves.

Although the results here do not establish any bounds on the minimum number of data points necessary for the QST-CGAN method to reconstruct a quantum state, they show that the QST-CGAN approach can perform much better than conventional reconstruction methods when data is scarce. An intuitive explanation of this is that since neural networks are universal function approximators, the \texttt{Generator} network might learn to find an approximation for the state in terms of a few parameters, e.g., the mean photon number for a thermal state, and estimate it better. However, the theoretical underpinnings of the QST-CGAN performance for few data points needs to be explored further, which is beyond the scope of this work.


\section{Conclusion}
\label{sec:conclusion}

We have shown how deep neural networks can assist in the characterization of quantum states. The states we considered here were optical quantum states, including bosonic error correction codes, but our methods are general and should be applicable also to systems with qubits.

\subsection{Classification}

We first showed how a neural network with convolutional layers followed by a dense layer can discriminate and classify several different types of quantum states with near-perfect accuracy. The input to the network was measurement data from phase-space descriptions of the states. The rare few misclassifications could be explained by the existence of parameter ranges where states from different classes are extremely similar and the problem of classification thus becomes ill-defined.

We further demonstrated the robustness of this classification method against two prominent noise sources --- additive Gaussian noise and single-photon loss. For the former, the network performance remained almost perfect until the standard deviation of the added noise reached as high values as \unit[20]{\%} of the largest input data values. For the latter noise, we showed a specific example where the network could identify a \texttt{cat} state even after it had lost \unit[70]{\%} of its initial photons.

By using the Grad-CAM method to extract and visualize which parts of the input phase space that the neural network bases its classification decision on, we proposed a simple adaptive technique for tomography that could significantly reduce the data-collection time for an experiment. Since the neural network learns the characteristic features of the states it is set to classify and can be trained to be robust against simple noise sources in the data, we can deploy it online at the initial stages of an experiment for guided data-collection during tomography.

\subsection{Reconstruction}

We next introduced, here and in Ref.~\cite{Ahmed2020}, a density-matrix-estimation technique using a combination of ideas from VAEs and GANs: the QST-CGAN method. This method uses custom neural-network layers that convert the output of any standard neural network into valid density matrices using the Cholesky decomposition. Therefore, we can convert any neural-network architecture into a variational map from input data to a density matrix. Following this scheme, we constructed a custom \texttt{Generator} network that maps input data to a density matrix and computes statistics for measurement operators.

By training the \texttt{Generator} network using gradient-based methods, we showed that the density matrix for the underlying state can be easily reconstructed. Instead of using a standard straight-forward loss function that requires an assumption on the likelihood for the data, we used a second \texttt{Discriminator} neural network to help train the \texttt{Generator}. Our choice of this adversarial training framework was motivated by an analysis of how standard loss functions, e.g., L2 or KL-divergence loss, perform for different states and noise in the data. We found that some of these standard loss functions resulted in good performance in the absence of noise, while other loss functions gave better performance in the presence of certain types of noise, but none of the loss functions led to a consistently good performance in a general setting. However, we showed that the QST-CGAN method is flexible and can easily adapt to different noise, states, or measurement settings. We ascribe this flexibility to the ability of the \texttt{Discriminator} to learn a loss function suited to the situation at hand.

We showed that the QST-CGAN-based reconstruction can be up to two orders of magnitude faster than iMLE, counted in the number of iterations required for reconstruction. Although the actual time for each iteration in the QST-CGAN can depend on the design of the neural networks, this presents a significant advantage for data post-processing during tomography. We also note that the neural-network based method seems to be performing non-trivial operations during reconstruction, e.g., applying a quantum operation to almost instantaneously jump from an orthogonal state to the correct state. This suggests that the neural networks learn to represent the state in a way that is well suited for the problem.

Having first benchmarked the reconstruction of pure states with no noise, we next considered how the QST-CGAN method can be augmented further to deal with noise in the data. We leveraged the flexibility of having a loss function that combines the $\texttt{Discriminator}$ loss with a simple L1 loss, since our objective is simply to make the generated data look like the training data. For the case of additive Gaussian noise of up to $5\%$ of the maximum signal value, our QST-CGAN method performs denoising and reconstruction much better than iMLE without needing any change in the architecture or loss function. Gaussian convolution noise corresponding to having a thermal state with mean photon number $n_{th}=5$ in a linear detection scheme was also tackled quite easily. The QST-CGAN only required the expected background noise as input which was added as special noise layers to the $\texttt{Generator}$ network.

Lastly, we showed that the QST-CGAN method clearly outperforms iMLE also when reconstructing mixed states. The QST-CGAN proved superior not only in terms of how few iterations it needed to reach high reconstruction fidelity, but also in terms of how little input data it required to reconstruct the state well. For a $\texttt{cat}$ state, the QST-CGAN required almost two orders of magnitude fewer data points than iMLE (as well as an RBM-based reconstruction shown in Ref.~\cite{Tiunov2020}) to achieve high reconstruction fidelity. It has been demonstrated that iMLE method can become stuck in cycles for some choices of input data, but our QST-CGAN method works well even with random sets of measurements generating the input data for the examples considered.

In conclusion, by connecting ideas of generative and discriminative modelling to quantum state classification and reconstruction, we have attempted to bridge the gap between deep neural networks and quantum information and computing. We have shown how some of the latest ideas from deep learning can be quite easily adapted and applied to quantum-information tasks with just a few tweaks to incorporate the rules of quantum physics. This opens up a wealth of possible applications, as we discuss further below and in Ref.~\cite{Ahmed2020}.


\section{Outlook}
\label{sec:outlook}

Our work suggests several new practical possibilities in data analysis of quantum experiments. At the same time, it leads to new questions regarding the limits of using neural networks for quantum state characterization.

It is expected that image recognition algorithms will be good at distinguishing different optical quantum states from their phase-space data. However, the benefit of using neural networks is their resilience to known types of noise. If we have to classify a ``cat'' state, it is rather easy to see that two lobes and a connecting bridge in the phase space should be a ``cat''. But what if, due to noise, the phase-space plots are shifted or rotated? An algorithm that relies on the fixed definition of a cat state will see poor overlap between the definition and the data, and hence cannot recognize the cat even if all the features are present. The neural-network method, on the other hand, is implicitly taught the important features that characterize cats and therefore works even in the presence of systematic or random noise.

In the case of reconstruction, we see that the the QST-CGAN method is a very powerful alternative to RBMs. We leverage the universal approximation capabilities of a deep neural network to have a tractable representation of the state by explicitly constructing the full density matrix. Standard loss functions such as fidelity, L1, L2, cross-entropy, etc.~will always have some shortcomings, since they require an assumption on the underlying likelihood for the data. Instead, with the CGAN framework, we let the loss metric be implicitly defined with the objective of simply making the data look similar to the generated data. However, we have not explored the theoretical underpinnings of using such a learned loss function for reconstruction. This remains to be analysed.

The future work that leverages these new ideas would go in two directions, beyond the suggestions for improvements and tweaks already mentioned in connection with the results. The first is further theoretical analysis of the techniques.  For example, it remains to be well understood how the neural network can reconstruct states using much fewer data points than a standard maximum likelihood method or perform non-trivial operations during a reconstruction. The second direction is validation with more experimental data and comparison to other standard methods for reconstruction. For example, we have not explored thoroughly how the QST-CGAN method compares with RBM-based approaches for tomography. This would be an interesting comparison since much of the work in QST with machine learning is focussed on using RBMs.

Since we ask for the full density matrix during reconstruction, our method cannot directly scale up for very large quantum systems. Even if it is straightforward to replace the density-matrix description with other efficient ans\"atze, it remains to be answered how to obtain efficient representations such that one does not use the millions of parameters in the deep neural network to estimate a few hundred parameters of the density matrix. 

The methods discussed here are ready to be applied to real experiments such that adaptive, online tomography schemes can be designed that can deal with noisy data. The techniques for classification and reconstruction could even be combined: the result of classifying a state with one neural network could be used to as a good starting point and parameterization for the training of another network for full quantum state reconstruction. We foresee that our ideas will lead to better techniques for quantum state characterization and bring the power of deep-learning-based tools to the quantum physicist.


\begin{acknowledgments}

We acknowledge useful discussions with Andreas Ekstr\"om, Giulia Ferrini, Simone Gasparinetti, Steven Girvin, Marina Kudra, Yong Lu, Florian Marquardt, Fernando Quijandr\'ia, and Ingrid Strandberg. We thank Yanming Che for comments on the manuscript. The neural networks were written using Tensorflow~\cite{Tensorflow2015}. Data generation and visualizations were done with QuTiP~\cite{Johansson2012, Johansson2013}, Scikit-image~\cite{scikit-image2014}, and Scikit-learn~\cite{Pedregosa2011}.

S.A.~and A.F.K.~acknowledge support from the Knut and Alice Wallenberg Foundation through the Wallenberg Centre for Quantum Technology (WACQT). 

C.S.M acknowledges that the project that gave rise to these results received the support of a fellowship from ``la Caixa'' Foundation (ID 100010434) and from the European Union's Horizon 2020 Research and Innovation Programme under the Marie Sk\l{}odowska-Curie Grant Agreement No.~847648. The fellowship code is LCF/BQ/PI20/11760026.

F.N. is supported in part by: NTT Research, Army Research Office (ARO) (Grant No.~W911NF-18-1-0358), Japan Science and Technology Agency (JST) (via the Q-LEAP program and CREST Grant No.~JPMJCR1676), Japan Society for the Promotion of Science (JSPS) (via the KAKENHI Grant No.~JP20H00134 and the JSPS-RFBR Grant No.~JPJSBP120194828), the Asian Office of Aerospace Research and Development (AOARD), and the Foundational Questions Institute Fund (FQXi) via Grant No.~FQXi-IAF19-06.

\end{acknowledgments}


\bibliography{references}

\end{document}